\documentclass[useAMS,usenatbib]{mn2e}
\usepackage{graphicx}
\usepackage{amssymb}

\title[GRB X-ray flares]{Unveiling the origin of X-ray flares in Gamma-Ray Bursts}
\author[G. Chincarini et al.]{G. Chincarini$^{1,11}$\thanks{E-mail: guido.chincarini@brera.inaf.it}, J. Mao$^{1,2,3}$, R. Margutti$^{1,11}$, M.G. Bernardini$^{1,4}$, C. Guidorzi$^{5}$, \and F. Pasotti$^{1}$, D. Giannios$^{6}$, M. Della Valle$^{7,4}$, A. Moretti$^{1}$, P. Romano$^{8}$, P. D'Avanzo$^{1}$, \and G. Cusumano$^{8}$ and P. Giommi$^{9,10}$\\
$^{1}$INAF - Osservatorio Astronomico di Brera, via Bianchi 46, I-23807 Merate (LC), Italy \\
$^{2}$Yunnan Observatory, Chinese Academy of Sciences, Kunming, Yunnan Province, 650011, China \\
$^{3}$Key Laboratory for the Structure and Evolution of Celestial Objects, Chinese Academy of Sciences, Kunming, \\
Yunnan Province, 650011, China \\
$^{4}$ICRANet, p.le della Repubblica 10, I-65100 Pescara, Italy\\
$^{5}$University of Ferrara, Physics Dept., via Saragat 1, I-44122 Ferrara, Italy\\
$^{6}$Department of Astrophysical Sciences, Peyton Hall, Princeton University, Princeton, NJ 08544, USA\\
$^{7}$INAF - Osservatorio di Capodimonte, Salita Moiariello 16, I-80131 Napoli, Italy\\
$^{8}$INAF - Istituto di Astrofisica Spaziale e Fisica Cosmica, Via La Malfa 153, I-90146 Palermo, Italy\\
$^{9}$ASI Science Data Center (ASDC), via G. Galilei, I-00044 Frascati (RM), Italy\\
$^{10}$ASI - Unit\'a Osservazione dell'Universo, viale Liegi 26, I-00198 Roma, Italy\\
$^{11}$University of Milano  Bicocca, Physics Dept., P.zza della Scienza 3, Milano 20126, Italy     }

\begin{document}

\date{Accepted 2010 Month day. Received 2010 Month day; in original form 2010 Month day}
\pagerange{\pageref{firstpage}--\pageref{lastpage}} \pubyear{.....}
\maketitle
\label{firstpage}

\begin{abstract}
We present an updated catalog of $113$ X-ray flares detected by \emph{Swift} in the  $\sim33\%$ of the  X-ray afterglows of Gamma-Ray Bursts (GRB). $43$ flares have a measured redshift. For the first time the analysis is performed in $4$ different X-ray energy bands, allowing us to constrain the evolution of the flare temporal properties with energy. We find that flares are  narrower at higher energies: their width follows a power-law relation $w\propto E^{-0.5}$ reminiscent of the prompt emission. Flares are asymmetric structures, with a decay time which is twice the rise time on average. Both time scales linearly evolve with time, giving rise to a constant rise-to-decay ratio: this implies that both time scales are stretched by the same factor.  As a consequence, the flare width \emph{linearly} evolves with time to larger values: this is a key point that clearly distinguishes the flare from the GRB prompt emission. The flare $0.3-10$ keV peak luminosity decreases with time, following a power-law behaviour with large scatter: $L_{pk}\propto t_{pk}^{-2.7\pm0.5}$. When multiple flares are present, a global softening trend is established: each flare is on average softer than the previous one. The $0.3-10$ keV isotropic energy distribution is a log-normal peaked at $10^{51}$ erg, with a possible excess at low energies. The flare average spectral energy distribution (SED) is found to be a power-law with spectral energy index $\beta\sim1.1$. These results confirmed that the flares are tightly linked to the prompt emission. However, after considering various models we conclude that no model is currently able to account for the entire set of observations.
\end{abstract}


\section{Introduction}\label{intro}

Gamma Ray Bursts (GRB) are short flashes of gamma-rays that during their early lifetime outshine any other source of gamma rays in the sky.  The first event was detected in 1967 and announced in 1973 \citep{1973ApJ...182L..85K}. Since then and after about 40 years of research our knowledge has increased significantly mainly thanks to three high-energy missions, \emph{Compton Gamma-Ray Observatory} (CGRO), \emph{Beppo}-SAX and \emph{Swift} that, together with related theoretical works, marked fundamental milestones in our knowledge of this phenomenon. These observations characterized the main features of these events. 

The timescale of the prompt emission lasts from a few milliseconds \citep{1981RSPTA.301..645V} to thousands of seconds \citep{2002astro.ph.11620H}. The distribution of its duration has been shown to be bimodal \citep{1981Ap&SS..80..119M,1984Natur.308..434N,1989cgrc.conf..337H,1992AIPC..265..304D,1993ApJ...413L.101K}, therefore GRBs can be classified as ``short'' and ``long''. The time profile of the prompt emission may present either multiple spikes of very short duration or relatively broad peaks with no fast variability \citep{2005ApJ...627..324N}. After the discovery of the isotropic distribution of the BATSE GRBs on the sky \citep{1992Natur.355..143M,1994ApJS...92..229F,1996ApJ...459...40B,1999ApJS..122..465P} and the detection of the afterglow of GRB970228 \citep{1997IAUC.6572....1C,1997IAUC.6584....1G,1997Natur.386..686V,1999ApJS..122..465P,2001ApJ...562..654D} it was clearly demonstrated that at least long GRBs were extragalactic and involved the emission of huge amounts of energy in a short time. After the launch of \emph{Swift} \citep{2004ApJ...611.1005G} it was firmly established that short GRBs also have an extragalactic origin \citep{2005Natur.437..851G,2005GCN..3570....1B,2005Natur.437..855V,2005Natur.438..994B} and therefore these bursts involve the emission of a rather large amount of energy as well. 

The X-Ray Telescope \citep[XRT][]{2005SSRv..120..165B} on board the \emph{Swift} satellite allows the early and well sampled observations of the afterglow. The temporal behaviour of the observed light curve was completely unexpected since, according to the data gathered by \emph{Beppo}-SAX (the record pointing of this satellite after detection was however of about $4$ hours), the expectation was for a flux decaying smoothly as a power-law, $F \sim t^{-1.5}$. It was realized rather soon by the \emph{Swift} team that the light curve of many GRBs was characterized by a more complex temporal behaviour \citep{2006ApJ...642..389N,2006ApJ...647.1213O,2006ApJ...642..354Z}: a steep early decay, a ``plateau'' and a late decay where the slope observed by \emph{Beppo}-SAX is essentially recovered. These phases can be either all or in part present \citep{2009MNRAS.397.1177E}. The significant achievements have been accompanied by substantial
theoretical effort to interpret the data. The internal - external shock model (\citealt{1993ApJ...418L..59M,1992MNRAS.258P..41R,1994ApJ...430L..93R}, see also \citealt{1999PhR...314..575P,2004RvMP...76.1143P} and references therein) within the fireball scenario \citep{1978MNRAS.183..359C} explains many of the characteristics of the observed light curve and spectrum of GRBs (see e.g. \citealt{2006ApJ...642..354Z} and reference therein; for a critical review of this model and possible alternatives see \citealt{2009arXiv0911.0349L}).

Indeed one of the most intriguing discoveries of the \emph{Swift}/XRT was the existence of flares in many of the observed GRB afterglows, that released a large emission of energy at later times than the prompt emission \citep{2005Sci...309.1833B,2006ApJ...641.1010F,2007ApJ...671.1903C,2007ApJ...671.1921F}. The first detection of flares with the \emph{Swift}/XRT occurred in X-Ray Flash (XRF) 050406 \citep{2005Sci...309.1833B,2006A&A...450...59R} and GRB050502B \citep{2005Sci...309.1833B,2006ApJ...641.1010F}. In the first case the afterglow light curve exhibits a rebrightening of a factor of $6$ that decayed quickly to recover the previous temporal behaviour. The flare in GRB050502B was spectacular with a rebrightening of the light curve of a factor $500$; its fluence is comparable to the one of the prompt emission observed by the Burst Alert Telescope \citep[BAT][]{2005SSRv..120..143B} on board the \emph{Swift} satellite. Further observations confirmed that flares are quite common events in the light curves of GRBs ($\sim 33\%$ of GRB afterglows exhibit flares). The energetics involved as well as their spectral properties, in particular the hard-to-soft evolution, are strong indications that X-ray flares have a common origin with the gamma-ray pulses. Furhermore, the presence of an underlying continuum with the same slope before and after the flaring activity excludes the possibility that flares are related to the afterglow emission by forward external shocks. Therefore their properties can provide an important clue toward the understanding of the mechanism that is at the basis of the GRB phenomenon.

Previous analysis was performed by \citet[][hereafter Paper I]{2007ApJ...671.1903C} and \citet[][hereafter Paper II]{2007ApJ...671.1921F}. These authors concluded that: 
\begin{enumerate}
\item	flares occur in all kind of GRBs: short and long, high energy peaked GRBs and X-ray flashes;
\item	the flare intensity decreases with time and the flare duration increases with time;
\item	a sizable fraction of flares cannot be related to the external shock mechanism;
\item	the temporal behaviour of flares is very similar to the one of prompt emission pulses;
\item	the number of flares of a single event does not correlate with the number of detected prompt pulses;
\item	the energy emitted during a bright flare is very large and in some cases is of the order of the prompt emission observed by BAT. Their average fluence, however, is about $10\%$ of the prompt emission fluence measured by BAT;
\item	the peak energy is typically in the soft X-rays, $\leq 1$ keV;
\item	the hardness ratio evolves following closely the evolution of the flare luminosity with a hardening during the rise and a softening during the decay;
\item	a long lasting activity by the central engine is advocated. 
\end{enumerate}

In the present work we expand the statistics of Paper I considering a wider sample of X-ray flares. Moreover, for the first time, we constrain the evolution of the properties of the flares in different X-ray energy bands inside the $0.3-10$ keV bandpass of the XRT. Finally, the flares are fitted with the function proposed in \citet{2005ApJ...627..324N}: this allows us to study the asymmetry of the flare temporal profiles, to assess the rise time and the decay time evolution with time.

The paper is organized as follows. In Sec. \ref{data_analysis} we describe the data reduction procedure and in Sec. \ref{sample} the flare sample and the fitting procedure. In Sec. \ref{analysis} we describe the analysis of the temporal behaviour of flares (Sec. \ref{shape}) and their energetic and spectral properties (Sec. \ref{energy}). In Sec. \ref{discussion} we discuss the main results of our analysis. Then conclusions follow.

Throughout the paper we follow the convention $f_{\nu}(t)\propto \nu^{-\beta}t^{-\alpha}$, where the energy spectral index $\beta$ is related to the photon index $\Gamma=\beta+1$. We have adopted the standard values of the cosmological parameters: $H_\circ=70$ km s$^{-1}$ Mpc$^{-1}$, $\Omega_M=0.27$ and $\Omega_{\Lambda}=0.73$. Errors are given at $1\, \sigma$ confidence level unless otherwise stated.

\section{Data reduction}\label{data_analysis}

XRT data were processed with the latest version of the \textsc{heasoft}
package available at the time of the discovery of the GRB explosion and corresponding
calibration files: standard filtering and screening criteria were applied.
In particular we used grades $0-2$ and $0-12$ in windowed timing (WT) and
photon counting (PC) modes, respectively.
\emph{Swift}/XRT \citep{2005SSRv..120..165B} is designed to acquire data using different observing modes
depending on source count rates to minimize the presence of pile-up
\citep{2004SPIE.5165..217H}: the first orbit contains Windowed Timing
(WT) data if the source is brighter than a few counts s$^{-1}$,
while for lower count rates the spacecraft automatically switches to the Photon Counting (PC) mode to
follow the fading of the source. Generally, WT data are extracted in a
$20\times 40$ pixel region centered at the afterglow position
along the WT strip of data. For count-rates above $\sim 150$ count s$^{-1}$,
we expect the source to be affected by pile-up \citep{2006A&A...456..917R}: in this case the central
part of the rectangular region is excluded from the analysis. The size of
the exclusion region is determined from the study of the distortion of the
grade $0$ distribution. The study of the distortion of the mean energy
associated to each photon gives consistent results.
WT background data are extracted within a rectangular box which is manually
chosen to be far from serendipitous background sources.

For PC observations, events are selected within four different regions: first, PC data
are extracted from a circular region centered at the enhanced position provided by
the XRT team. In most cases a radius of $20$ pixel ($1$ pixel $\sim2.36''$)
is used. The radius is chosen so as to contain $\sim 90\%$ of the  total flux as determined
by the \textsc{xrtmkarf} tool. Exceptions are however present: bright (faint)
sources require  radii greater (smaller) than the standard $20$ pixel value;
at the same time, also for bright sources, a region of event extraction smaller than usual
is sometimes necessary  to avoid contamination from serendipitous background sources.
When the PC data suffered from pile-up, we extracted the source events in an annulus
whose inner radius  is derived comparing the observed to the nominal Point Spread Function \citep[PSF,][]{2006ApJ...638..920V}.
The original radial distribution of radiation -and the total source flux- is then recovered
using the accurate information of the instrumental PSF provided by
\citet{2005SPIE.5898..360M}. When the count-rate of the fading afterglow is lower than
$0.01$ count s$^{-1}$, the source events were extracted from a smaller region to assure
a high signal-to-noise (SN) ratio. This required the introduction of a third PC
extraction region: a circle with a typical  $10$ pixel radius.
Finally, the PC background level was assessed extracting the events from an annular region
centered as close as possible to the GRB location and with
an inner radius greater than $3/2$ the radius of the source region. When this was not possible,
background data were selected within a circular region with size as big as possible. In both cases
the background region is chosen in a source-free portion of the soft X-ray sky.
This was accomplished by reading the PC cleaned event list file with the \textsc{Ximage}
package and localizing any source in the field with a minimum SN equal to $2$.

The background subtracted, PSF and vignetting corrected light-curves were then re-binned so as
to assure a minimum SN equals to $4$ for WT and PC mode data. When single-orbit data were not
able to fulfill the signal-to-noise requirement, data coming from different orbits are merged
to build a unique data point. Further details can be found in Margutti et al. (in prep.).

This procedure was applied $5$ times to produce a $0.3-10$ keV
count-rate light curve (``tot'') together with four count-rate light curves of the same event
in $4$ different sub-energy bands. In particular we have:
``fvs'': light curve containing photons filtered in the observed energy interval $0.3-1$ keV;
``fms'': light curve containing photons filtered in the observed energy interval $1-2$ keV;
``fmh'': light curve containing photons filtered in the observed energy interval $2-3$ keV;
``fvh'': light curve containing photons filtered in the observed energy interval $3-10$ keV.
This assures the valuable possibility to investigate the variations of the flare temporal
properties as a function of the energy of the emitted photons.

\section{Sample Selection and fitting procedure}\label{sample}

We considered all the \emph{Swift} GRBs observed between April 2005 and March 2008. During this period the GRBs detected were $332$ ($109$ with redshift $z$): $284$ by \emph{Swift} ($104$ with $z$), $26$ by INTEGRAL ($1$ with $z$), $10$ by IPN ($1$ with $z$), $9$ by HETE ($3$ with $z$), and $3$ by AGILE. Observations of GRB afterglows by \emph{Swift}/XRT have been carried out in $234$ cases\footnote{For the numbers of observed GRBs provided we refer to http://www.mpe.mpg.de/~jcg/grbgen.html}.
 
In order to have a sample as homogeneous as possible, we selected flares from all the X-ray afterglows observed by XRT with the following criteria:
\begin{enumerate}
\item the flare contains a relatively complete structure: rise, peak and decay phase; 
\item the flare structure can be fitted with an analytic function, thus giving a homogeneous set of parameters; 
\item the flare is clearly distinguishable from the underlying continuum. Small fluctuations have not been identified as flares; 
\item blended flares, which have more complicated structures, are included in the present statistical sample only if the various pulses are easily identified;
\item the flares are ``early'', i.e. peak time $<1000$ s.
\end{enumerate}

The present data set consists of $56$ GRBs that have bright (i.e. peak count rate $>1$ count s$^{-1}$) X-ray flares. Two of them, GRB051210 and GRB070724A, are short GRBs. $27$ GRBs present a single flare, $15$ two flares and in $14$ cases we have more than two flares. The total number of flares is $113$, each one analysed in the XRT energy bandpass ($0.3-10$ keV) and, when the count-rate was high enough to properly fit the profile of the emission ($30$ flares), also in sub-energy bands ($0.3-1$ keV, $1-2$ keV, $2-3$ keV, $3-10$ keV) in order to detect any chromatic difference. $21$ GRBs of the present sample have also redshift measurements ($43$ flares): this subsample has been used to investigate the flare properties in the source rest frame.

Some giant flares, such the ones detected in GRB060510B, GRB070129 and GRB070616, are not included: the initial flare in GRB060510B could be the prompt emission observed in X-rays, and the underlying continuum is hard to define; flares in GRB071029 are strongly overlapping; the flare in GRB070616 contains some fluctuations which are difficult to distinguish.

The choice of limiting the present sample only to early flares is motivated by the aim of making a direct comparison with the prompt emission pulses. Therefore, the early flares, with higher SN ratio and temporal resolution, are more suitable for this purpose. A detailed analysis of late flares and a comparison with this sample will be presented in Bernardini et al. (in prep.).

\subsection{Fitting procedure}\label{fit}

The flares detected by XRT generally present a rather smooth profile and only in very few cases it is possible to detect substructure indicating rapid variability, although to a rather low statistical level. Previous attempts to detect variability over very small time scales have been inconclusive. This subject, beyond the scope of the present work, will be discussed in Margutti et al. (in prep.).

To derive information about the properties of the flares, and eventually their relation to the shock physical parameters, two approaches are viable: a) to develop a model based on the assumed physical mechanism or b) to choose an empirical analytical function to fit the observed light curve and to estimate the parameters characterizing its shape. In this second case the comparison with models prescriptions can be carried out later. We choose this second empirical approach since it allows to obtain more general results that should then help in discriminating among different models and to avoid biases. 

In Paper I we used a Gaussian and broken power-law functions for the combined fit flare plus underlying continuum to derive widths and peak intensities. For the rise and decaying times and slopes we used an exponential and, for a few cases, we tested the function introduced in \citet{1997ApJ...490...92K}. 

In this work we derive the flare parameters using the function proposed by \citet[][Norris05]{2005ApJ...627..324N}, after testing various empirical or semi-empirical profiles. The function used in \citet{2005ApJ...627..324N} consists of two combined exponentials defined by $5$ parameters that is not differentiable at the peak time. Nevertheless it is quite flexible to reproduce a wide range of pulse shapes and width. \citet{1997ApJ...490...92K} (but see also \citealt{1998MNRAS.296..275D}) proposed a semi-empirical profile that would account for the variability produced by internal shocks among relativistic shells having an ad hoc distribution of the Lorentz factor. Although such an issue is beyond the scope of the present work, the function proposed is quite satisfactory. This profile is not differentiable at the peak intensity as well, since it is the sum of two functions that are used to fit the rising and decaying part separately. \citet{2003ApJ...596..389K} proposed a FRED profile using an analytic function derived from physical first principles and accounting for the spectral evolution. The derived function describes the pulses by $4$ parameters and the peak time can be calculated from them. The profile is completely satisfactory and leads to clear statistical results for single pulses, as it will be discussed later on. On the other hand, when compared to the Norris05 profile it appears to be slightly more complicated and the derivation of the parameters of interest less forthcoming. For all these reasons we choose the Norris05 profile. 

The Norris05 function is the inverse of the product of two exponentials and is fully determined by $4$ parameters:
\begin{equation}
\rm{Norris05}(t)= A\, \lambda \, e^{-\frac{\tau_1}{(t-t_s)}-\frac{(t-t_s)}{\tau_2}}\, \, \, \rm{for}\, \, t>t_s
\end{equation}
where $\mu=(\tau_1/\tau_2)^{1/2}$ and $\lambda=e^{2\mu}$. The intensity is at maximum at $t_{pk}=\tau_{pk}+t_s=(\tau_1 \tau_2)^{1/2}+t_s$. The pulse width is measured between the two $1/e$ intensity points,
\begin{equation}
w=\Delta t_{1/e}=t_{\rm decay}+t_{\rm rise}=\tau_2 (1+4\mu)^{1/2}\, .
\end{equation}
The pulse asymmetry is
\begin{equation}
k=\frac{t_{\rm decay}-t_{\rm rise}}{t_{\rm decay}+t_{\rm rise}}=(1+4\mu)^{-1/2}\, .
\end{equation}
$t_{\rm decay}$ and $t_{\rm rise}$ are expressed in terms of $w$ and $k$ as:
\begin{equation}
t_{\rm decay,rise}=\frac{1}{2}w(1\pm k)\,.
\end{equation}
For a full description of the Norris05 function and parameters we refer to \citet{2005ApJ...627..324N}. The results of the fitting procedure are listed in Tables \ref{tab_flares} and \ref{tab_bkg}. The errors on the derived quantities are obtained accounting for the entire covariance matrix of the fitting parameters. The flare fluence $S$ is calculated integrating the corresponding Norris05 function over the interval between $t_{1,90}$ and $t_{2,90}$, and then converted into physical units by applying a conversion factor computed by extracting the spectrum around the peak time with a minimum of $2000$ photons. The spectrum has been fit using an absorbed simple power-law model within XSPEC.

\begin{figure}
\includegraphics[width=\hsize,clip]{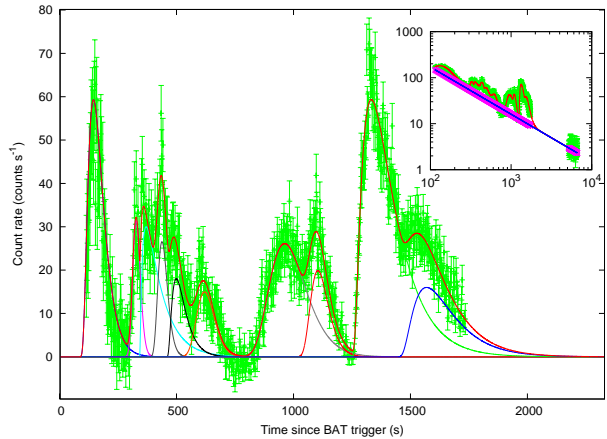}
\caption{The best-fitting of the flares of GRB051117A with $11$ Norris05 function (colored lines) is shown after the subtraction of the underlying continuum. \emph{Inset:} Simultaneous fit of the same flares (red solid line) and of the underlying continuum (blue solid line). The errors on the estimate of the underlying continuum are properly accounted for (pink error bars).}
\label{flare_fit}
\end{figure}

We fitted simultaneously the underlying continuum and the flares by adopting a multiply broken power-law plus a number of Norris05 functions (the best-fitting of the flares of GRB051117A is shown in Fig. \ref{flare_fit} as an example). As a measure of the quality of the fit we used the $\chi^2$ statistics. Quite often, and especially for the $0.3-1$ keV and $3-10$ keV bandpasses, the fits were not satisfactory due to the low count-rate and SN ratio. However the fit could be mildly constrained by the solution of the parameters we had when fitting the whole XRT band ($0.3-10$ keV).

\begin{table*}
\tiny
 \begin{minipage}{180mm}
  \caption{Table of the best-fitting parameters and of the derived physical quantities. From left to right: name of the GRB, bandpass (see Sec. \ref{data_analysis}), redshift $z$, parameters describing the Norris05 function ($A$, $t_s$, $\tau_1$, $\tau_2$), peak time $t_{pk}$, width $w$, asymmetry $k$, rise time $t_{\rm rise}$, decay time $t_{\rm decay}$, fluence of the flare $S$.}
  \label{tab_flares}
    \resizebox{\textwidth}{!}{
\begin{tabular}{lllllllllllll}
\hline
\hline
GRB   & band  & $z$	&   $A$          &  $t_s$               &   $\tau_1$             & $\tau_2$               & $t_{pk}$              &  $w$                   &     $k$               &   $t_{\rm rise}$       	   & $t_{\rm decay}$	     	          & $S$  \\
        &	&	&   count s$^{-1}$     	  & s	  	      &s	    	    &s	                  &s                    &s	  	      &                     &   s       	   &s	                  &$10^{-8}$erg cm$^{-2}$   \\
\hline  																	        		      
050406  & tot   &--	&$  4.6\pm  0.4  $&$107.0 \pm	90.8 $&$  903.5 \pm 2404.0 $&$   11.7 \pm    9.6 $&$  209.6 \pm    4.0 $&$   70.1 \pm	  7.1$&$   0.2  \pm    0.1 $&$   29.2 \pm      6.4$&$	40.9 \pm      5.5$&$ 0.9\pm   0.6$ \\ 
050406  & fvs	&--	&$  2.7\pm  0.5  $&$ 160.0\pm	 59.4$&$   429.2\pm  1341.0$&$    8.0 \pm     7.9$&$   218.6\pm     5.5$&$    44.0\pm	  7.6$&$    0.2 \pm	0.2$&$    18.0\pm      5.2$&$	 26.0\pm      5.7$&$ 0.2\pm   0.2$  \\  
050406  & fms	&--	&$  1.5\pm  0.3  $&$ 169.2\pm	 15.5$&$    36.0\pm    71.0$&$   22.6 \pm    12.1$&$   197.8\pm     6.5$&$    55.6\pm	 10.5$&$    0.4 \pm	0.2$&$    16.5\pm      6.8$&$	 39.1\pm      9.1$&$ 0.1\pm   0.0$  \\
050406  & fmh	&--	&$   --\pm   --  $&$	--\pm	   --$&$      --\pm	 --$&$      --\pm      --$&$	  --\pm      --$&$	--\pm	   --$&$     -- \pm	 --$&$      --\pm	--$&$	   --\pm       --$&$ 0.0\pm   0.0$  \\
050406  & fvh	&--	&$  0.7\pm  4.8  $&$ 180.5\pm	 27.9$&$     0.4\pm    38.0$&$    40.0\pm    97.4$&$   184.3\pm   166.1$&$    47.1\pm	253.5$&$    0.8 \pm	6.2$&$     3.5\pm    164.3$&$	 43.6\pm     99.3$&$ 0.1\pm   0.6$  \\
050502B & tot1  &--	&$ 14.6\pm  3.7  $&$ 350.8\pm	 39.4$&$   656.5\pm   381.5$&$   141.6\pm    27.0$&$   655.7\pm   101.2$&$   438.9\pm	 86.6$&$    0.3 \pm	0.0$&$   148.7\pm     34.3$&$	290.2\pm     54.2$&$24.0\pm   7.7$  \\
\hline	
\end{tabular}	}
\end{minipage}
\end{table*}	 	 	 
			 	 	 
\setcounter{table}{0}
\begin{table*}
\tiny
 \begin{minipage}{180mm}
  \caption{Continued}
    \resizebox{\textwidth}{!}{
  \begin{tabular}{lllllllllllll}
\hline
\hline
GRB   & band  & $z$	&   $A$          &  $t_s$               &   $\tau_1$             & $\tau_2$               & $t_{pk}$              &  $w$                   &     $k$               &   $t_{rise}$       	   & $t_{decay}$	     	          & $S$  \\
        &	&	&   count s$^{-1}$     	  & s	  	      &s	    	    &s	                  &s                    &s	  	      &                     &   s       	   &s	                  &$10^{-8}$erg cm$^{-2}$   \\
\hline																					  
051117A & fmh4  &  -- &$    3.2\pm     5.0  $&$  24.9\pm  3807.0$&$132105.0\pm 4.01E06$&$     1.3\pm	15.1$&$   435.4\pm	--$&$	 45.8\pm    61.1$&$	0.0\pm     0.3$&$    22.3\pm	23.3$&$    23.5\pm    37.9$&$	0.4\pm   2.4$ \\
051117A & fmh5  &  -- &$    2.2\pm     1.1  $&$ 453.6\pm    55.4$&$    50.8\pm   254.6$&$    37.3\pm   101.4$&$   497.1\pm    15.5$&$	 88.8\pm   112.8$&$	0.4\pm     0.7$&$    25.7\pm	17.9$&$    63.0\pm   105.8$&$	0.5\pm   1.1$ \\
051117A & fmh6  &  -- &$    1.3\pm     3.8  $&$-439.2\pm  8219.0$&$324995.0\pm 8.60E06$&$     3.4\pm	37.1$&$   612.7\pm	--$&$	119.7\pm   196.5$&$	0.0\pm     0.3$&$    58.2\pm	80.4$&$    61.6\pm   116.3$&$	0.4\pm   1.3$ \\
051117A & fmh7  & --  &$    3.3\pm     0.3  $&$ 584.1\pm   478.1$&$  6358.8\pm 26200.0$&$    21.4\pm	32.4$&$   953.4\pm    20.9$&$	179.2\pm    31.8$&$	0.1\pm     0.2$&$    78.9\pm	11.6$&$   100.3\pm    29.9$&$	1.4\pm   1.6$ \\
051117A & fmh8  &  -- &$    2.1\pm     1.4  $&$ 674.7\pm  2579.0$&$ 38928.0\pm 654500.0$&$     4.7\pm	21.7$&$  1100.6\pm	--$&$	 89.1\pm    67.0$&$	0.1\pm     0.3$&$    42.2\pm	43.5$&$    46.9\pm    24.2$&$	0.4\pm   1.3$ \\
051117A & fmh9  & --  &$    7.4\pm     0.4  $&$1234.4\pm    15.0$&$   132.4\pm    79.1$&$    68.3\pm	14.6$&$  1329.5\pm     5.0$&$	175.1\pm    11.9$&$	0.4\pm     0.1$&$    53.4\pm	 5.0$&$   121.7\pm    12.4$&$	4.4\pm   1.0$ \\
\hline
\end{tabular}}
\end{minipage}																	
\end{table*}

\setcounter{table}{0}
\begin{table*}
\tiny
 \begin{minipage}{180mm}
  \caption{Continued}
    \resizebox{\textwidth}{!}{
  \begin{tabular}{lllllllllllll}
\hline
\hline
GRB   & band  & $z$	&   $A$          &  $t_s$               &   $\tau_1$             & $\tau_2$               & $t_{pk}$              &  $w$                   &     $k$               &   $t_{rise}$       	   & $t_{decay}$	     	          & $S$  \\
        &	&	&   count s$^{-1}$     	  & s	  	      &s	    	    &s	                  &s                    &s	  	      &                     &   s       	   &s	                  &$10^{-8}$erg cm$^{-2}$   \\
\hline
060707  & fmh	&3.425  &$     --\pm	  --  $&$    --\pm	--$&$	   --\pm      --$&$	 --\pm      --$&$      --\pm	  --$ &$      --\pm	 --$&$      --\pm      --$&$	  --\pm      --$&$	--\pm	   --$&$    0.0\pm    0.0$\\
060707  & fvh	&3.425  &$    2.2\pm	54.1  $&$ 152.9\pm  3154.0$&$	 18.3\pm  2585.0$&$    31.5\pm   144.5$&$   176.9\pm  1521.0$ &$    63.4\pm  1460.9$&$     0.5\pm    13.7$&$	15.9\pm   800.0$&$    47.5\pm	661.5$&$    1.0\pm   15.0$\\
060714  & tot1  &2.711  &$  118.7\pm	 4.6  $&$ 103.9\pm     3.2$&$	  3.0\pm     2.8$&$    33.5\pm     1.9$&$   113.9\pm	 1.5$ &$    49.6\pm	4.7$&$     0.7\pm     0.1$&$	 8.1\pm     3.0$&$    41.6\pm	  2.0$&$   33.0\pm    4.3$\\
060714  & tot2  &2.711  &$   89.6\pm	12.1  $&$  55.6\pm    75.7$&$ 26926.1\pm 54480.0$&$	0.2\pm     0.0$&$   132.1\pm	 1.8$ &$     8.2\pm	4.1$&$     0.0\pm     0.0$&$	 4.0\pm     2.1$&$     4.2\pm	  2.1$&$    3.8\pm    1.3$\\ 
060714  & tot3  &2.711  &$  121.8\pm	13.6  $&$ 131.5\pm     4.2$&$	 12.8\pm    12.3$&$	5.2\pm     1.0$&$   139.7\pm	 1.4$ &$    14.0\pm	1.9$&$     0.4\pm     0.1$&$	 4.4\pm     1.1$&$     9.6\pm	  1.0$&$    9.5\pm    2.4$\\
060714  & tot4  &2.711  &$  120.6\pm	 4.9  $&$ 151.5\pm     4.7$&$	127.6\pm    76.1$&$	4.5\pm     0.8$&$   175.4\pm	 0.5$ &$    21.1\pm	0.8$&$     0.2\pm     0.0$&$	 8.3\pm     0.6$&$    12.8\pm	  0.5$&$   11.0\pm    2.0$\\
\hline
\end{tabular}}
\end{minipage}																	
\end{table*}

\setcounter{table}{0}
\begin{table*}
\tiny
 \begin{minipage}{180mm}
  \caption{Continued}
    \resizebox{\textwidth}{!}{
  \begin{tabular}{lllllllllllll}
\hline
\hline
GRB   & band  & $z$	&   $A$          &  $t_s$               &   $\tau_1$             & $\tau_2$               & $t_{pk}$              &  $w$                   &     $k$               &   $t_{rise}$       	   & $t_{decay}$	     	          & $S$  \\
        &	&	&   count s$^{-1}$     	  & s	  	      &s	    	    &s	                  &s                    &s	  	      &                     &   s       	   &s	                  &$10^{-8}$erg cm$^{-2}$   \\
\hline
070518  & tot1  &--	&$   23.2\pm	 2.1  $&$  97.8\pm     2.1$&$	  3.5\pm     4.3$&$    16.7\pm     5.3$&$   105.4\pm	 5.3$&$    28.2\pm     9.3$&$	  0.6\pm     0.1$&$	5.7\pm     3.0$&$    22.5\pm	 7.0$&$      2.6\pm    0.8$\\
070518  & tot2  &--	&$    7.2\pm	 1.3  $&$ 130.9\pm     5.4$&$	  8.6\pm    13.4$&$    36.0\pm    37.0$&$   148.5\pm	17.3$&$    61.9\pm    55.5$&$	  0.6\pm     0.2$&$    13.0\pm    11.3$&$    48.9\pm	45.8$&$      1.8\pm    1.5$\\
070518  & tot3  &--	&$   22.2\pm	 1.7  $&$ 151.8\pm    11.5$&$	179.1\pm   174.3$&$	7.6\pm     2.4$&$   188.6\pm	22.1$&$    34.2\pm    11.6$&$	  0.2\pm     0.1$&$    13.3\pm     5.0$&$    20.9\pm	 6.7$&$      2.9\pm    0.9$\\
070518  & fvs1  &--	&$    6.9\pm	 1.0  $&$  59.2\pm    84.0$&$	669.8\pm  3299.0$&$	4.3\pm     7.4$&$   112.7\pm   163.0$&$    30.6\pm    54.4$&$	  0.1\pm     0.2$&$    13.1\pm    24.6$&$    17.4\pm	30.0$&$      0.8\pm    1.0$\\
070518  & fvs2  &--	&$    5.8\pm	 6.9  $&$ 133.8\pm     4.2$&$	  0.1\pm     3.8$&$    14.3\pm    17.7$&$   135.2\pm	19.5$&$    16.9\pm    37.6$&$	  0.8\pm     1.6$&$	1.3\pm    16.1$&$    15.6\pm	24.6$&$      0.4\pm    0.7$\\
070518  & fvs3  &--	&$   11.7\pm	 0.9  $&$ 105.1\pm    76.4$&$  1255.3\pm  3372.0$&$	5.6\pm     4.8$&$   189.2\pm   141.1$&$    43.9\pm    40.5$&$	  0.1\pm     0.1$&$    19.1\pm    18.6$&$    24.8\pm	22.0$&$      1.8\pm    1.3$\\
\hline			        												       
\end{tabular}		 }
\end{minipage}		        													      
\end{table*}
\begin{table*}
\tiny
\begin{minipage}{180mm}
  \caption{Table of the best-fitting parameters and of the derived physical quantities of the underlying continuum. From left to right: name of the GRB, bandpass (see Sec. \ref{data_analysis}), parameters describing the fitting function (power law or broken power law): normalization $N$, power law indices $\alpha_1$ and $\alpha_2$, break time $t_{br}$, boundaries of the flare $T_{90}$, $\chi^2/dof$ of the simultaneous fit of the flares and of the underlying continuum.}
  \label{tab_bkg}
\begin{tabular}{lllllllllllll}
\hline
\hline
GRB & band & N               & $\alpha_1$ & $\alpha_2$ & $t_{br}$ & $t_{1,90}$ & $t_{2,90}$ & $\chi^2/dof$\\
    &	     & count s$^{-1}$	 &	         	&	           & s  	    & s          & s	        &	   \\
\hline
050406  & tot   &$ 229.3  \pm 160.0  $&$     1.1\pm	0.1$&$      --\pm      --$&$	  --\pm      -- $& 174.9  & 262.0  &	0.7\\
050406  & fvs	&$   664.9\pm	808.2$&$     1.4\pm	0.2$&$      --\pm      --$&$	  --\pm      -- $&   196.5&   254.0&	1.3\\
050406  & fms	&$   257.1\pm	439.9$&$     1.4\pm	0.3$&$      --\pm      --$&$	  --\pm      -- $&   177.9&   262.7&	0.8\\
050406  & fmh	&$    --  \pm	   --$&$      --\pm	 --$&$      --\pm      --$&$	  --\pm      -- $&	--&	 --&	 --\\
050406  & fvh	&$   151.1\pm	265.9$&$     1.4\pm	0.3$&$      --\pm      --$&$	  --\pm      -- $&   180.6&   296.2&	0.9\\
050502B & tot1  &$    42.0\pm	 34.0$&$     0.9\pm	0.1$&$      --\pm      --$&$	  --\pm      -- $&   479.4&  1073.9&	1.2\\
\hline
\end{tabular}
\end{minipage}																
\end{table*}

\setcounter{table}{1}
\begin{table*}
\tiny
\begin{minipage}{180mm}
  \caption{Continued}
\begin{tabular}{lllllllllllll}
\hline
\hline
GRB & band & N               & $\alpha_1$ & $\alpha_2$ & $t_{br}$ & $t_{1,90}$ & $t_{2,90}$ & $\chi^2/dof$\\
    &	     & count s$^{-1}$	 &	         	&	           & s  	    & s          & s	        &	   \\
\hline
050922B & fmh3  &$1.00E10 \pm2.55E10 $&$     3.7\pm	0.4$&	 --&	--&  1018.4&  1152.6&	  1.1 \\
050922B & fvh1  &$4.00E11 \pm1.19E12 $&$     4.3\pm	0.5$&	 --&	--&   652.0&   661.0&	  0.8\\ 
050922B & fvh2  &$4.00E11 \pm1.19E12 $&$     4.3\pm	0.5$&	 --&	--&   645.9&  1091.4&	  0.8\\ 
050922B & fvh3  &$4.00E11 \pm1.19E12 $&$     4.3\pm	0.5$&	 --&	--&   927.0&  1001.1&	  0.8\\ 
051117A & tot1  &$ 18983.1\pm  2563.0$&$     1.0\pm	0.0$&	 --&	--&   106.8&   231.3&	1.0 \\	
051117A & tot2  &$ 18983.1\pm  2563.0$&$     1.0\pm	0.0$&	 --&	--&   303.3&   358.5&	1.0 \\
\hline
\end{tabular}
\end{minipage}
\end{table*}

\setcounter{table}{1}
\begin{table*}
\tiny
\begin{minipage}{180mm}
  \caption{Continued}
\begin{tabular}{lllllllllllll}
\hline
\hline
GRB & band & N               & $\alpha_1$ & $\alpha_2$ & $t_{br}$ & $t_{1,90}$ & $t_{2,90}$ & $\chi^2/dof$\\
    &	     & count s$^{-1}$	 &	         	&	           & s  	    & s          & s	        &	   \\
\hline	                                                                                                  
060210  & fvs1 &$1.50E06 \pm2.89E06 $&$     2.4\pm     0.4$&$	  0.2\pm     0.5$&$   215.9\pm    47.3 $& 168.1&   269.6  &	1.5\\ 060210  & fvs2 &$1.50E06 \pm2.89E06 $&$     2.4\pm     0.4$&$	  0.2\pm     0.5$&$   215.9\pm    47.3 $& 340.0&   475.6  &	1.5\\	
060210  & fms1 &$4.01E07 \pm6.34E07 $&$     3.0\pm     0.3$&$	  0.8\pm     0.2$&$   192.9\pm    21.2 $& 169.9&   245.9  &	1.7\\	
060210  & fms2 &$4.01E07 \pm6.34E07 $&$     3.0\pm     0.3$&$	  0.8\pm     0.2$&$   192.9\pm    21.2 $& 356.0&   430.2  &	1.7\\	
060210  & fmh1 &$1.50E08 \pm3.59E08 $&$     3.5\pm     0.5$&$	  0.9\pm     0.4$&$   205.5\pm    31.0 $& 169.3&   237.2  &	1.3\\	
060210  & fmh2 &$1.50E08 \pm3.59E08 $&$     3.5\pm     0.5$&$	  0.9\pm     0.4$&$   205.5\pm    31.0 $& 358.3&   417.5  &	1.3\\
\hline	          											  
\end{tabular}
\end{minipage}     																	   
\end{table*}

\setcounter{table}{1}
\begin{table*}
\tiny
\begin{minipage}{180mm}
  \caption{Continued}
\begin{tabular}{lllllllllllll}
\hline
\hline
GRB & band & N               & $\alpha_1$ & $\alpha_2$ & $t_{br}$ & $t_{1,90}$ & $t_{2,90}$ & $\chi^2/dof$\\
    &	     & count s$^{-1}$	 &	         	&	           & s  	    & s          & s	        &	   \\
\hline
060729  & fmh	&$  1.0E29\pm 3.88E29$&$    12.6\pm	0.8$&$   3.7\pm   1.9$&$ 188.4\pm  18.8 $&   166.9&   199.3    &     1.1 \\
060729  & fvh	&$	--\pm	   --$&$      --\pm	 --$&$    --\pm    --$&$    --\pm    -- $&	--&	 --    &      -- \\
060804  & tot	&$     7.0\pm	 17.6$&$     4.8\pm	0.5$&$   0.0\pm    --$&$    --\pm    -- $&   422.2&   662.1    &     0.8 \\
060804  & fvs	&$	--\pm	   --$&$      --\pm	 --$&$    --\pm    --$&$    --\pm    -- $&	--&	 --    &      -- \\
060804  & fms	&$	--\pm	   --$&$      --\pm	 --$&$    --\pm    --$&$    --\pm    -- $&	--&	 --    &      -- \\
060804  & fmh	&$     1.3\pm	  0.8$&$     0.4\pm	0.1$&$    --\pm    --$&$    --\pm    -- $&   389.1&   583.8    &     0.6 \\
\hline
\end{tabular}
\end{minipage}																
\end{table*}

\setcounter{table}{1}
\begin{table*}
\tiny
\begin{minipage}{180mm}
  \caption{Continued}
\begin{tabular}{lllllllllllll}
\hline
\hline
GRB & band & N               & $\alpha_1$ & $\alpha_2$ & $t_{br}$ & $t_{1,90}$ & $t_{2,90}$ & $\chi^2/dof$\\
    &	     & count s$^{-1}$	 &	         	&	           & s  	    & s          & s	        &	   \\
\hline
070419B & fvh  &$200070.0\pm 89830.0$&$     1.8\pm     0.1$&	--	--&    --    -- &   199.1&   306.0&	2.3\\
070518  & tot1 &$  5000.0\pm  6861.0$&$     1.3\pm     0.3$&	--	--&    --    -- &    99.0&   143.5&	0.7\\
070518  & tot2 &$  5000.0\pm  6861.0$&$     1.3\pm     0.3$&	--	--&    --    -- &   133.9&   233.9&	0.7\\
070518  & tot3 &$  5000.0\pm  6861.0$&$     1.3\pm     0.3$&	--	--&    --    -- &   173.1&   215.4&	0.7\\
070518  & fvs1 &$   103.6\pm   110.6$&$     0.7\pm     0.2$&	--	--&    --    -- &    97.0&   134.9&	0.8\\
070518  & fvs2 &$   103.6\pm   110.6$&$     0.7\pm     0.2$&	--	--&    --    -- &   133.8&   171.7&	0.8\\
\hline
\end{tabular}
\end{minipage}																
\end{table*}	

\setcounter{table}{1}
\begin{table*}
\tiny
\begin{minipage}{180mm}
  \caption{Continued}
\begin{tabular}{lllllllllllll}
\hline
\hline
GRB & band & N               & $\alpha_1$ & $\alpha_2$ & $t_{br}$ & $t_{1,90}$ & $t_{2,90}$ & $\chi^2/dof$\\
    &	     & count s$^{-1}$	 &	         	&	           & s  	    & s          & s	        &	   \\
\hline																			    									    
080212	&fvs3	&$ 3.08E12\pm 5.98E12$&$     5.3\pm	0.4$&$     1.1\pm      0.1$&$	200.1\pm     14.5$ &   274.6&	392.6&     1.1\\
080212  & fms1  &$ 7.01E12\pm 1.09E13$&$     5.4\pm	0.3$&$     1.1\pm      0.0$&$	182.4\pm      7.6$ &   176.4&	266.9&     1.7\\
080212  & fms2  &$ 7.01E12\pm 1.09E13$&$     5.4\pm	0.3$&$     1.1\pm      0.0$&$	182.4\pm      7.6$ &   219.9&	309.0&     1.7\\
080212  & fms3  &$ 7.01E12\pm 1.09E13$&$     5.4\pm	0.3$&$     1.1\pm      0.0$&$	182.4\pm      7.6$ &   273.9&	353.3&     1.7\\
080212  & fmh1  &$ 7.01E12\pm 3.00E13$&$     5.8\pm	0.9$&$     1.1\pm      0.0$&$	157.5\pm      8.3$ &   181.6&	233.9&     1.0\\
080212  & fmh2  &$ 7.01E12\pm 3.00E13$&$     5.8\pm	0.9$&$     1.1\pm      0.0$&$	157.5\pm      8.3$ &   218.7&	306.2&     1.0\\
\hline
\end{tabular}
\end{minipage}																	
\end{table*}

\section{Analysis}\label{analysis}

\subsection{Flares shape parameters}\label{shape}

In the following we revisit with this new sample the results obtained in Paper I and Paper II. In addition to a more detailed analysis of the flares temporal behaviour, the major contribution of the present work is the analysis of the flares in $4$ XRT sub-energy bandpasses (see Sec. \ref{data_analysis}). When not specified, we consider the parameters of the flares obtained fitting the $0.3-10$ keV band light curve. The phenomenology of the flares is presented in the observer frame unless otherwise stated.

\subsubsection{Width versus $t_{pk}$}\label{wtpk}

\begin{figure}
\includegraphics[width=\hsize,clip]{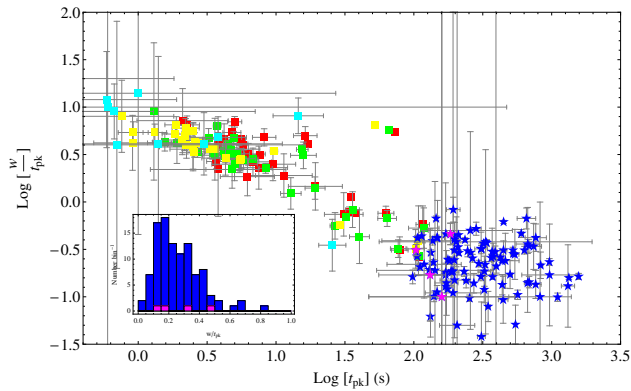}
\caption{Plot of the ratio $w/t_{pk}$ of X-ray flares (blue and pink dots correspond to flare belonging to long and short GRBs respectively) compared with the one of the BATSE pulses in 4 sub-energy bandpasses in the \citet{2005ApJ...627..324N} sample (colored dots). Their different behaviour is evident: while the BATSE pulse ratio $w/t_{pk}$ decreases with time in all the energy bands, for the flares it remains constant up to late times. \emph{Inset:} distribution of the ratio $w/t_{pk}$ of X-ray flares (blue and pink rectangles correspond to flare belonging to long and short GRBs respectively). The distribution is centered on $w/t_{pk}=0.23$ with
$\sigma=0.14$}
\label{WTOTPK_BATSE_XRT}
\end{figure}

The first property of flares that we want to investigate is their temporal variability. Based on the fits described in Sec. \ref{fit}, we calculate the ratio between the width and the peak time: $w/t_{pk}$. The median value of the distribution is $0.23$ and the standard deviation is $0.14$ (see Fig. \ref{WTOTPK_BATSE_XRT}, Inset). This value is in agreement with the result of Paper I, since the width obtained with the Norris05 profile corresponds to the one measured at $\sim 37\%$ of the maximum, and if we consider the width measured at $37\%$ of the maximum on a Gaussian profile, it results to be $w_{\rm Gauss} = 2.83\, \sigma_{\rm Gauss}$. 

\begin{figure}
\includegraphics[width=\hsize,clip]{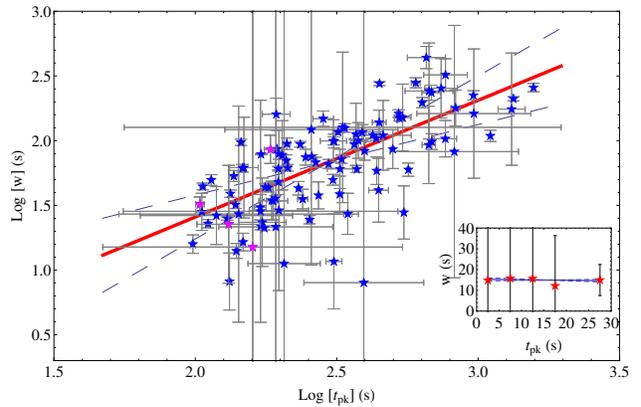}
\caption{Plot of the observed width as a function of the observed peak time for the sample of the X-ray flares (blue and pink dots correspond to flare belonging to long and short GRBs respectively). Red solid line: best-fitting $w=10^{(-0.4\pm0.9)}\,t_{pk}^{(0.9\pm0.4)}$. \emph{Inset:} median of the width of pulses with peak time within different intervals for BATSE sample from \citet{2005ApJ...627..324N} (red points), showing that $w$ remains constant with time during the prompt emission.}
\label{WIDTH_PEAK}
\end{figure}

What is evident in Fig. \ref{WIDTH_PEAK} is that the flare width increases linearly with time with a best fit $w\sim 0.2\,t_{pk}$. This correlation is quite strong, with a Spearman rank coefficient $\rho=0.7115$ (number of points $N=109$, null hypothesis probability $\rm{nhp}<2.2\times 10^{-16}$), and it is confirmed also in the source rest frame, thus excluding an effect due to the redshift. This is remarkably different from what has been found for the prompt emission pulses observed by BATSE: the width remains constant throughout the GRB time history \citep[][see also Fig. \ref{WIDTH_PEAK}, inset]{2000ApJ...539..712R}. By comparing the BATSE broad pulses from the \citet{2005ApJ...627..324N} sample with X-ray flares, this different trend is evident: while the $w/t_{pk}$ of the BATSE pulses decreases with time in all the 4 sub-energy bandpasses, for XRT flares it tends to be constant up to late times (see Fig. \ref{WTOTPK_BATSE_XRT}). It is important to note, however, that apart from the different trend the ratio $w/t_{pk}$ of the flares joints smoothly with the one observed in the BATSE broad pulses, thus pointing to a continuous transition from one class to the other.

\subsubsection{Width versus Energy}

\begin{figure}
\includegraphics[width=\hsize,clip]{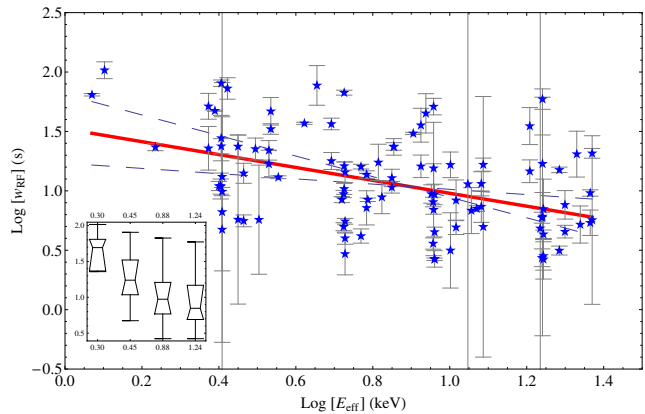}
\caption{Plot of the rest frame width as a function of the rest frame effective energy for the X-ray flares fitted in all the 4 sub-energy bandpasses (blue dots). Red solid line: best-fitting $w_{\rm RF}=10^{(1.5\pm0.3)}\, E_{\rm eff}^{(-0.5\pm0.3)}$. \emph{Inset:} Boxplot of the same quantities.}
\label{WIDTH_EN}
\end{figure}

We considered the sub-sample of flares that have been fit in all the 4 sub-energy bandpasses that have also a redshift measurement to investigate the existence of a correlation between width and energy band. This analysis has been carried out in the source rest frame. We defined therefore the effective rest frame energy in the observed energy band $[h\nu_1;h\nu_2]$ as:
\begin{equation}
E_{\rm eff}=(1+z)\frac{\int_{h\nu_1}^{h\nu_2} h\nu\, f(\nu)\, \rm{RM}(\nu)\, d(h\nu)}{\int_{h\nu_1}^{h\nu_2} f(\nu)\, \rm{RM}(\nu)\, d(h\nu)}\, ,
\end{equation}
where $\rm{RM}(\nu)$ is the response matrix of the XRT and $f(\nu)$ is the spectrum, $f(\nu)\propto \nu^{-\beta}$ with $\beta\simeq 1$. By correlating the width measured in the source rest frame ($w_{\rm RF}=w/(1+z)$) with the rest frame effective energy of each sub-energy band we find: $w_{\rm RF}\propto E_{\rm eff}^{-0.5}$ (see Fig \ref{WIDTH_EN}). We conclude that flares are broader at lower energies. It is interesting to observe, at this point, that while flares are very clearly detected at X-ray frequencies they are almost undetectable over the optical light curve (see e.g. \citealt{2009ApJ...697..758K}, however with the exception of the optical flare of GRB080129, \citealt{2009ApJ...693.1912G}). Whether or not this is a consequence of the present correlation it is unknown and a new analysis is in progress to establish that. 

\citet{1995ApJ...448L.101F} reported in the analysis of BATSE pulses an inverse correlation between the width of a peak and the energy band in which the observation is made, $w \propto E^{-0.4}$, confirmed later on by \citet{1996ApJ...459..393N}, and by \citet{2007A&A...465..765B} that extended this analysis also to the prompt emission pulses detected by the \emph{Beppo}–SAX satellite in the X-ray and gamma-ray bands. 

\subsubsection{Asymmetry and rise and decay times}

\begin{figure}
\includegraphics[width=\hsize,clip]{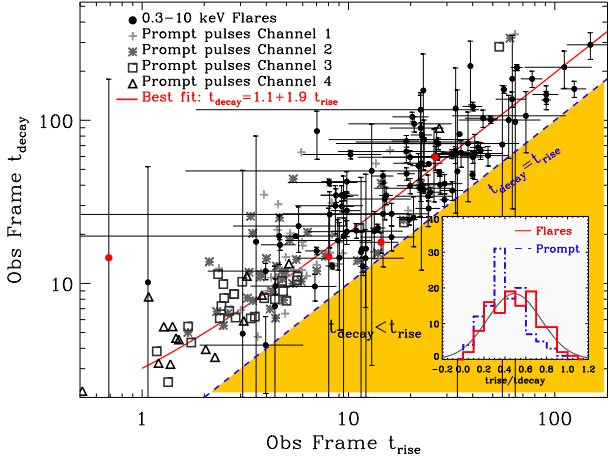}
\caption{Observed decay time as a function of the rise time
for our sample of X-ray flares (black and red dots correspond to flare belonging to long and short GRBs respectively) and the sample of prompt pulses
from \citet{2005ApJ...627..324N}: the different symbols refer to different BATSE
energy bandpasses. Red solid line: best linear fitting for the two samples:
$t_{\rm decay}=(1.1\pm0.1)+(1.9\pm0.1)\,t_{\rm rise}$. Blue dashed line:
$t_{\rm decay}=t_{\rm rise}$ locus of points. Shaded area: region of the plane for
which $t_{\rm decay}<t_{\rm rise}$. No data point lies in this region.
\emph{Inset:} distribution of the $t_{\rm rise}/t_{\rm decay}$  values for the
sample of X-ray flares (red solid line) and the sample of prompt pulses
of \citet[][blue dot-dashed line]{2005ApJ...627..324N}.
The best Gaussian fitting of the flare distribution is also shown with a solid
grey line. The distribution is centered on $t_{\rm rise}/t_{\rm decay}=0.49$ with
$\sigma=0.26$. }
\label{Obs_trise_vs_tdecayPAPER}
\end{figure}

Another characterizing property of the flare temporal behaviour is their asymmetry. Flares are asymmetric with a median value of $0.35$ and standard deviation of $0.2$. Furthermore, the asymmetry $k$ does not 
correlate with the width $w$, thus indicating that flares profiles are 
extremely self-similar. As a direct comparison, the distribution of asymmetry values from the 
$4$ bandpass BATSE sample of \citet{2005ApJ...627..324N} is characterized by a median
of $0.49$ and standard deviation of $0.26$. The two distributions are portrayed in
the inset of Fig. \ref{Obs_trise_vs_tdecayPAPER}. \citet{2003ApJ...596..389K} found a distribution of 
$t_{\rm rise}/t_{\rm decay}$ values centered on $0.47$ for a sample of separable BATSE 
pulses\footnote{Note, however, that in \citet{2003ApJ...596..389K} the rise and decay times 
were defined at the profile half maximum.}, in good agreement with our
results. Flares and prompt pulses share very similar values of asymmetry.
Despite their similarity, the flare distribution of $t_{\rm rise}/t_{\rm decay}$
is more symmetric while the prompt pulses distribution is skewed to lower
values. This feature can be understood considering the results that will be discussed in detail in Margutti et al. (in prep.): these authors found that both the
evolution of the rise and the evolution of the decay times with energy band
can be described by a power-law, with a steeper decaying power-law index
associated to the rise time evolution. This means that in a softer energy
band (like the X-ray when compared to the prompt gamma-ray photons) the pulses
will be allowed to have a larger $t_{\rm rise}/t_{\rm decay}$ ratio.

Fig. \ref{Obs_trise_vs_tdecayPAPER} further demonstrates the presence of a 
strong correlation between the rise and decay times: $t_{\rm decay}\propto t_{\rm rise}$, with $\rho=0.7856$ ($N=113$, $\rm{nhp}<2.2\times 10^{-16}$).
The correlation smoothly joins the prompt and the flare emission, from the
gamma-ray to the soft X-rays, from the trigger to thousands of seconds after 
the onset of the explosion. No emission episode in these two samples has
$t_{\rm decay}<t_{\rm rise}$.  

A trend is seen between the rise and decay times as a function of the 
flare peak time, reminiscent of the width-peak time correlation of Fig.
\ref{WIDTH_PEAK}: since the rise time is proportional to the decay time and 
$w=t_{\rm rise}+t_{\rm decay}$ by definition, the results portrayed in Fig.
\ref{Obs_trise_vs_tdecayPAPER} naturally account for the rise and decay times vs. peak time linear
correlations: $t_{\rm rise}\sim 0.05\,t_{pk}$ and $t_{\rm decay}\sim 0.14\,t_{pk}$.

\subsection{Energy and spectrum}\label{energy}

\subsubsection{Intensity versus $t_{pk}$}

In Paper I we detected an inverse correlation between the flare intensity $A$ and $t_{pk}$. Moreover, it as been shown in the previous section that the flare duration $w$ increases with $t_{pk}$. Therefore, flares seem to have a tendency to be equally energetic, lasting longer when the intensity is much lower. 

\begin{figure}
\includegraphics[width=\hsize,clip]{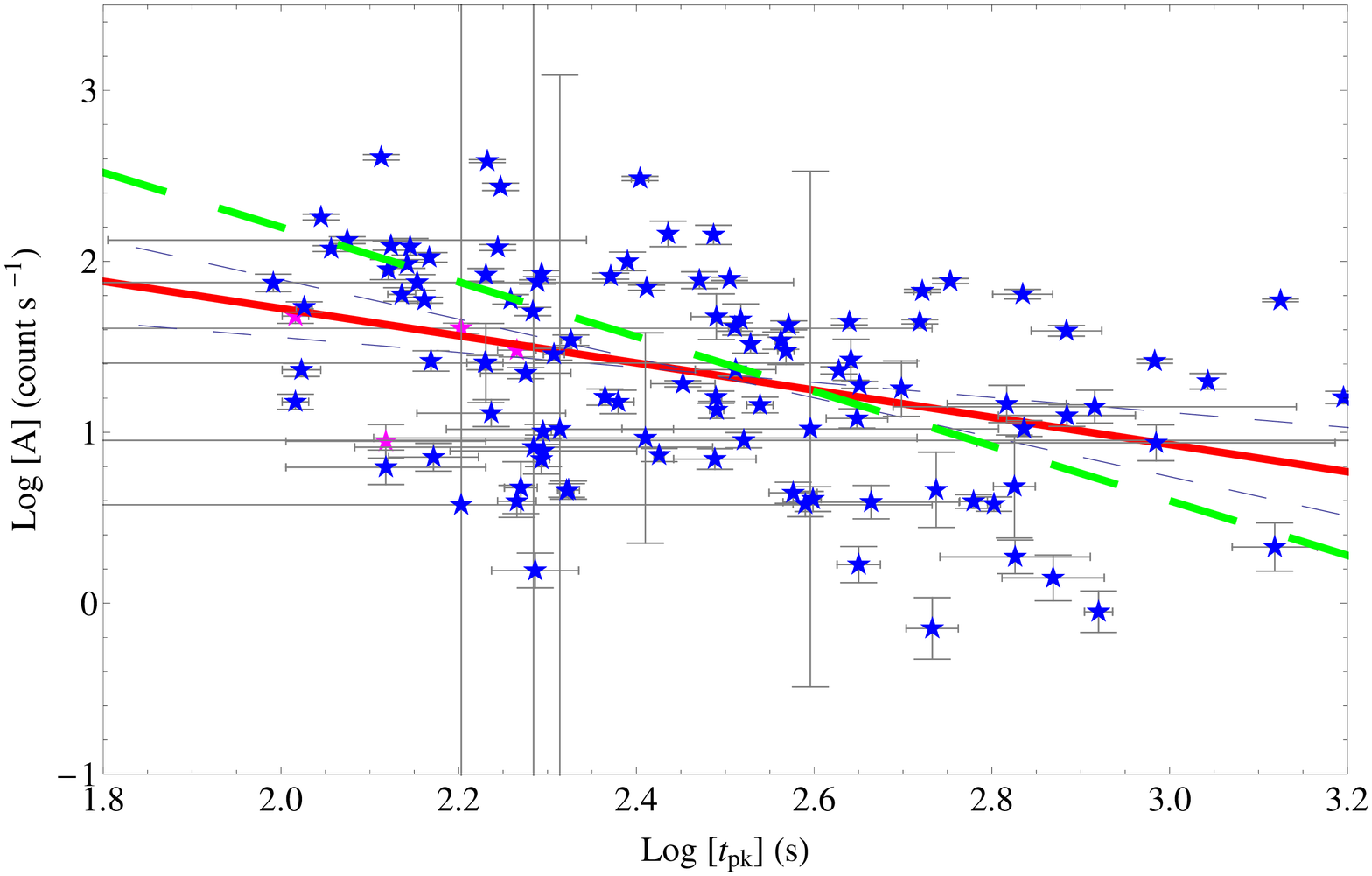}
\includegraphics[width=\hsize,clip]{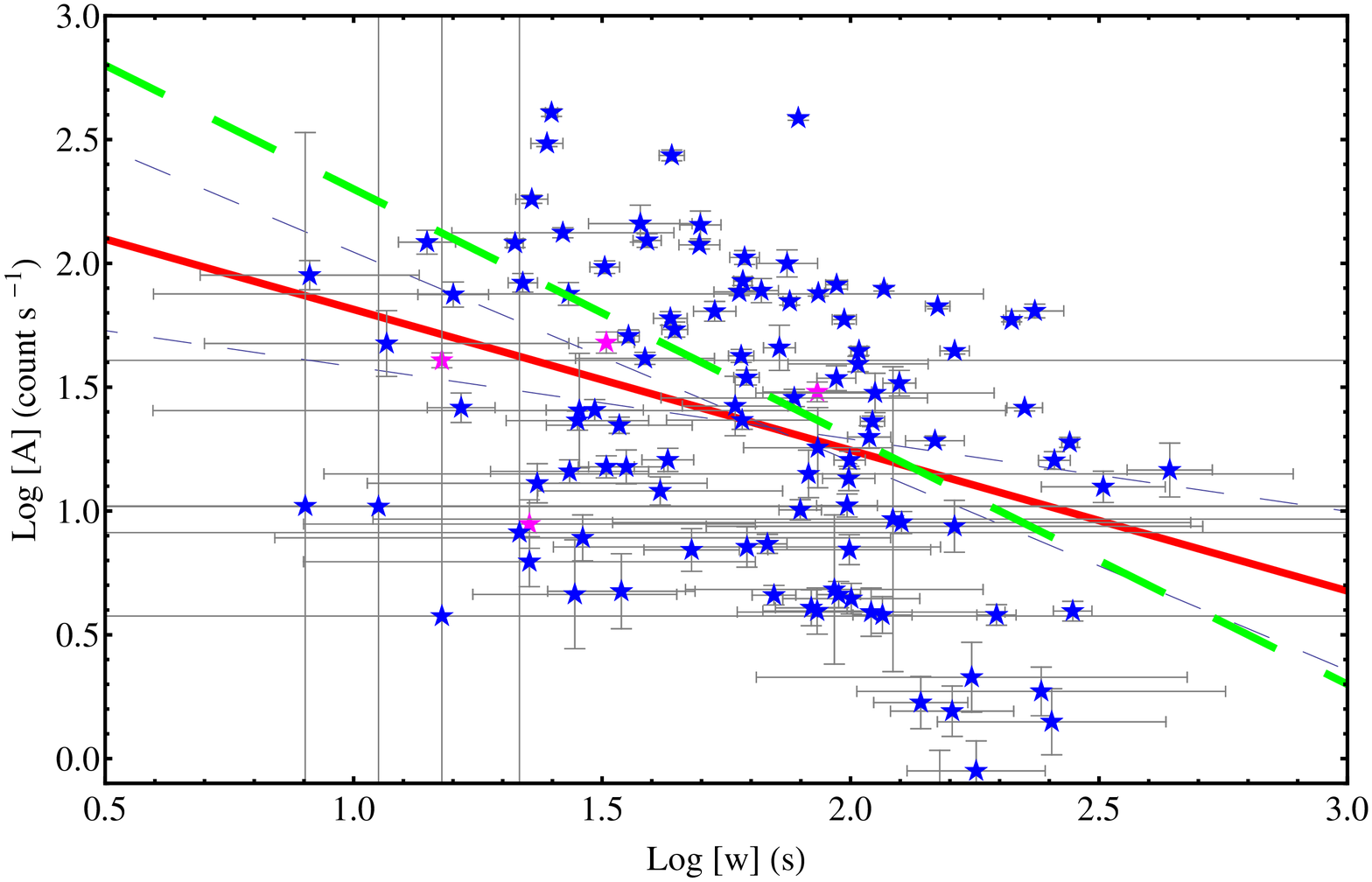}
\caption{\emph{Upper panel:} Plot of the peak intensity $A$ of the X-ray flares as a function of the peak time $t_{pk}$ (blue and pink dots correspond to flares belonging to long and short GRBs respectively). Red solid line: best-fitting $A=10^{(3.3\pm0.9)}\,t_{pk}^{(-0.70\pm0.01)}$. Green dashed line: best-fitting obtained accounting for the intrinsic dispersion of the data \citep[further details can be found in][]{2005physics..11182D}: $Log(A)=q+m\, Log(t_{pk})$, with $q=(5.4\pm0.8)$, $m=(-1.6\pm0.4)$ and the extrinsic scatter $\sigma=(0.55\pm0.07)$.
\emph{Lower panel:} Plot of the peak intensity $A$ of the X-ray flares as a function of the width $w$ (blue and pink dots correspond to flare belonging to long and short GRBs respectively). Red solid line: best-fitting $A=10^{(2.3\pm0.5)}\,w^{(-0.6\pm0.3)}$. Green dashed line: best-fitting obtained accounting for the intrinsic dispersion of the data \citep[further details can be found in][]{2005physics..11182D}: $Log(A)=q+m\, Log(w)$, with $q=(3.3\pm0.5)$, $m=(-1.0\pm0.3)$ and the extrinsic scatter $\sigma=(0.55\pm0.07)$.}
\label{NORM_WIDTH_TPEAK}
\end{figure}

The previous sample could cast some doubts on the correlation intensity peak time since it was heavily weighted on a few very late flares. For the early flares of that sample, the resulting correlation was not statistically significant. The present sample is restricted to flares occurring earlier than $\sim 1000$ s in the observer frame ($\leq 320$ s in the rest frame) where the statistics are higher. The correlation between these two quantities\footnote{The flare peak intensity is derived as a best fitting parameter of the Norris05 profile and is defined as the excess over the underlying continuum.} is shown in Fig. \ref{NORM_WIDTH_TPEAK}, upper panel: despite the rather large errors, the correlation is evident with $\rho=-0.3921$ ($N=109$, $\rm{nhp}=2.9\times 10^{-5}$). 

A correlation also exists between $A$ and the width $w$ with $\rho=-0.3442$ ($N=109$, $\rm{nhp}=2.7\times 10^{-4}$; see Fig. \ref{NORM_WIDTH_TPEAK}, lower panel). We find that $A\propto w^{-0.6}$. This correlation has the same trend to that of the prompt emission pulses \citep{2000ApJ...539..712R} but the the power-law index for the prompt emission is remarkably different ($A\propto w^{-2.8}$).

\begin{figure}
\includegraphics[width=\hsize,clip]{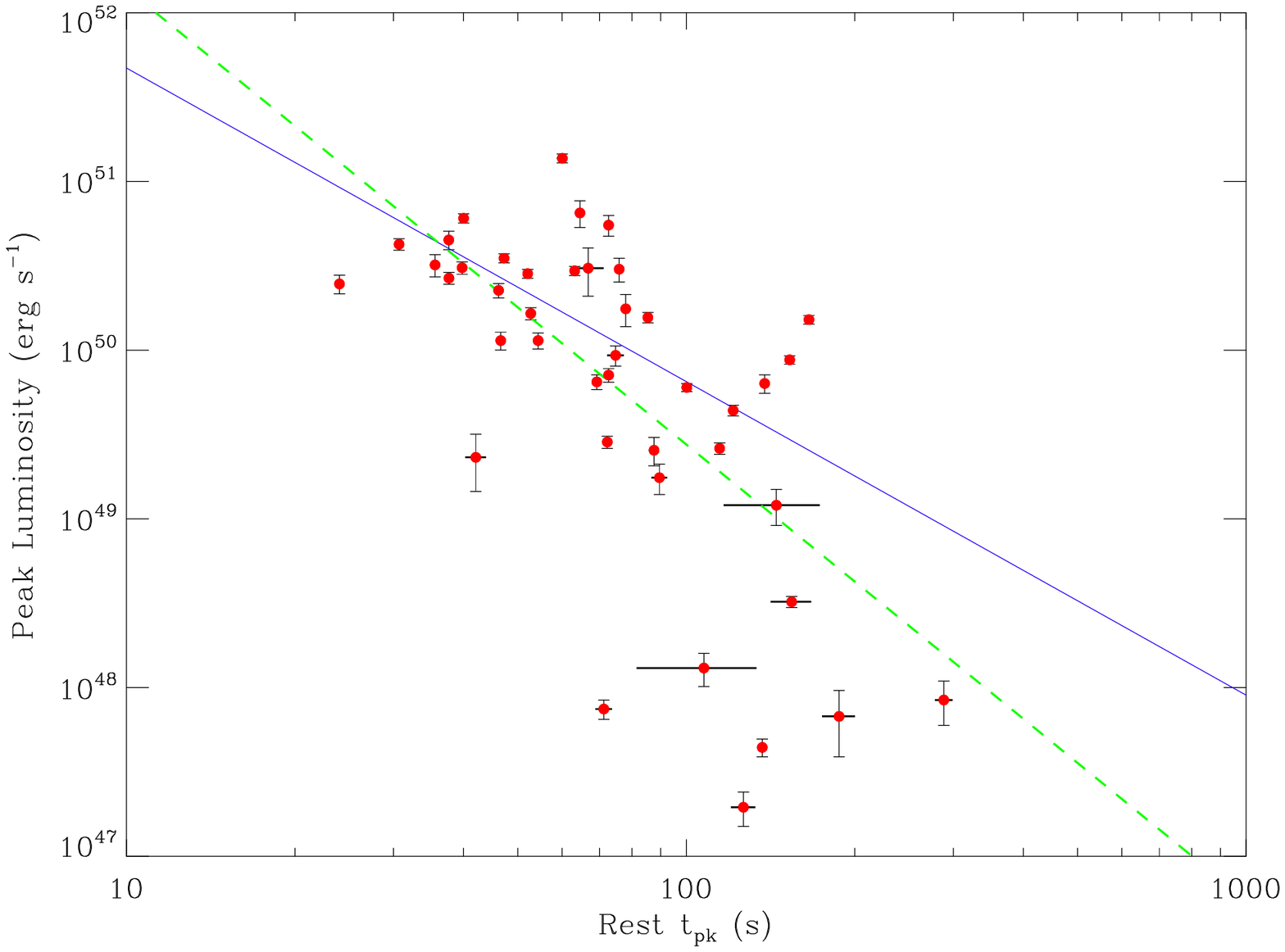}
\includegraphics[width=\hsize,clip]{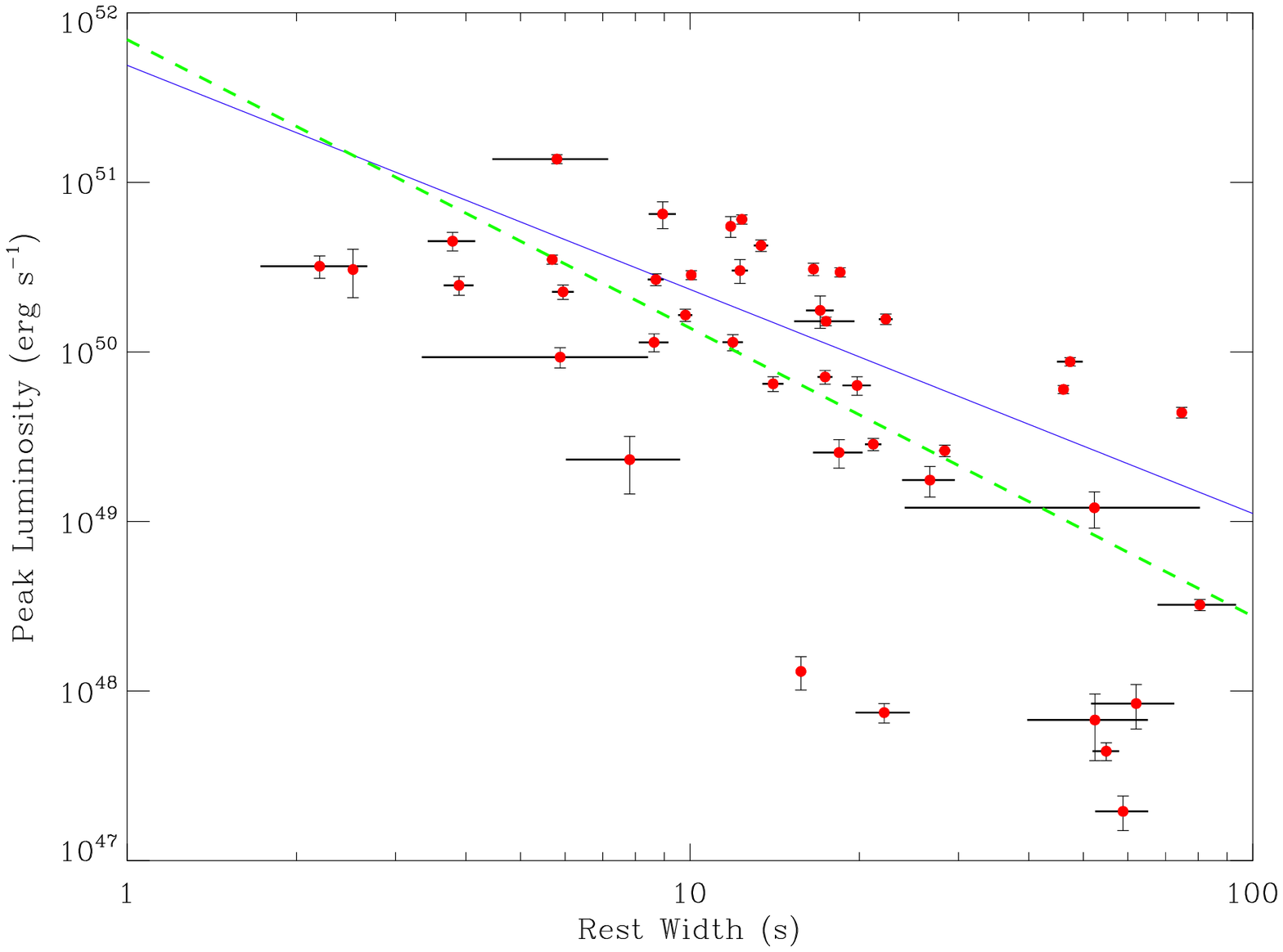}
\caption{\emph{Upper panel:} Peak luminosity vs rest frame peak time for the subsample of flares
with measured redshift. The blue solid line indicates
the best-fitting relation:
$L_{pk}=10^{(49.9\pm0.1)}\, t_{pk,\rm{RF}}^{(-1.9\pm0.1)}$. The green dashed line indicates the best-fitting obtained accounting for the intrinsic dispersion of the data \citep[further details can be found in][]{2005physics..11182D}: $Log(L_{pk})=q+m\, Log(t_{pk,\rm{RF}})$, with $q=(54.8\pm4.0)$, $m=(-2.7\pm0.5)$ and the extrinsic scatter $\sigma=(0.73\pm0.08)$.
\emph{Lower panel:} Peak luminosity vs rest frame width for the subsample of flares
with measured redshift. The blue solid line indicates
the best-fitting power-law relation: $L_{pk}=10^{(48.1\pm0.1)}\,w_{\rm RF}^{(-1.3\pm0.1)}$. The green dashed line indicates the best-fitting obtained accounting for the intrinsic dispersion of the data \citep[further details can be found in][]{2005physics..11182D}: $Log(L_{pk})=q+m\, Log(w_{\rm RF})$, with $q=(51.8\pm3.9)$, $m=(-1.7\pm0.3)$ and the extrinsic scatter $\sigma=(0.68\pm0.08)$.}
\label{PeakLum_vs_tpeak}
\end{figure}

The existence of a correlation between the flare peak intensity
and the flare peak time is confirmed in the rest frame plane, as well.
Fig. \ref{PeakLum_vs_tpeak} shows the existence of a trend between the peak
luminosity and the flare rest frame peak time: flares occurring later are
characterized by  lower peak luminosity. The best fit relation is found
to be: $L_{pk}\propto t_{pk}^{-1.9}$. The Spearman rank coefficient is $\rho=-0.7240$ ($N=43$, $\rm{nhp}=6.0\times 10^{-8}$). 
The best-fitting relation is dominated by the group of bright flares for
which the parameters are determined with the highest accuracy:
however, Fig. \ref{PeakLum_vs_tpeak} clearly show the presence of a group of low
luminosity flares ($L_{pk}<10^{45}$ erg s$^{-1}$) which lies
below the the prediction and would indicate a much steeper relation. If we properly account for the presence of intrinsic scatter \citep{2005physics..11182D}, the best-fitting relation reads: $Log(L_{pk})=q+m\, Log(t_{pk,\rm{RF}})$, with $q=(54.8\pm4.0)$, $m=(-2.7\pm0.5)$. We note that this relation describes the evolution of the peak
luminosity of single flares: the flare \emph{mean} luminosity
function can evolve differently with time. For comparison, \citet{2008MNRAS.388L..15L}
found that the mean luminosity, averaged over a time-scale longer than
the duration of the individual flares, declines as a power-law in time
with index $\sim1.5$.    

The existence of the peak time vs. width correlation of Sec. \ref{wtpk}
together with the result of Fig. \ref{PeakLum_vs_tpeak}, upper panel, automatically translates into
a peak luminosity vs. rest frame width relation portrayed in Fig. \ref{PeakLum_vs_tpeak}, lower panel.
The best fit power-law model reads: $L_{pk}\propto t_{pk}^{-1.3}$ and $\rho=-0.6871$ ($N=43$, $\rm{nhp}=5.0\times 10^{-7}$). Accounting for the intrinsic scatter we have: $Log(L_{pk})=q+m\, Log(w_{\rm RF})$, with $q=(51.8\pm3.9)$, $m=(-1.7\pm0.3)$.
Flares becomes wider and less luminous as the time proceeds.

\subsubsection{Energy}

\begin{figure}
\includegraphics[width=\hsize,clip]{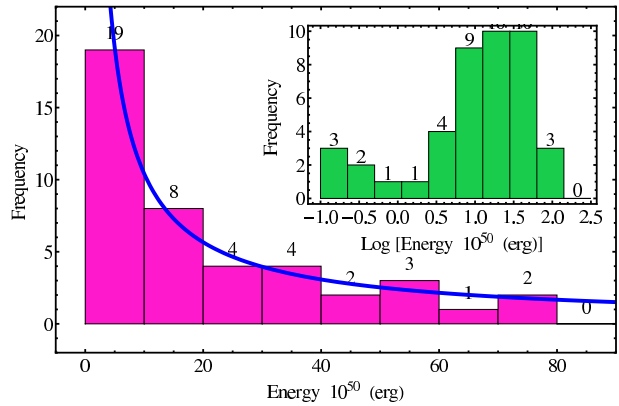}
\caption{Distribution of the energy emitted in single X-ray flares of the present sample $E_{\rm flare}$. \emph{Inset:} Distribution of the logarithm of the energy to better evidence the faint tail of the distribution.}
\label{ENERGY_HISTO}
\end{figure}

The isotropic $0.3-10$ keV energy of the X-ray flares of the sample can be estimated from the fluence $S$ as:
\begin{equation}
E_{\rm flare} = \frac{4\pi D_l^2}{(1+z)}S\, .
\end{equation}
The distribution function of the energy emitted in each flare cannot be properly determined since the sample is not complete and homogeneous. However, from Fig. \ref{ENERGY_HISTO} we see that the average isotropic energy emitted in single X-ray flares is about $10^{51}$ erg, $5$ to $10\%$ of the energy observed by BAT in the prompt emission. Moreover, there is a hint of a bimodal distribution of the logarithm of the energy of flares, thus revealing a possible excess of faint flares.

At high redshift we have bright flares that we do not detect at lower redshift, however, under the assumption that bright flares correlate with bright GRBs, this could be due to the smaller volume sampled at low redshift. Likewise the brightest flares occur at earlier times (peak time $< 500$ s). 

\begin{figure}
\includegraphics[width=\hsize,clip]{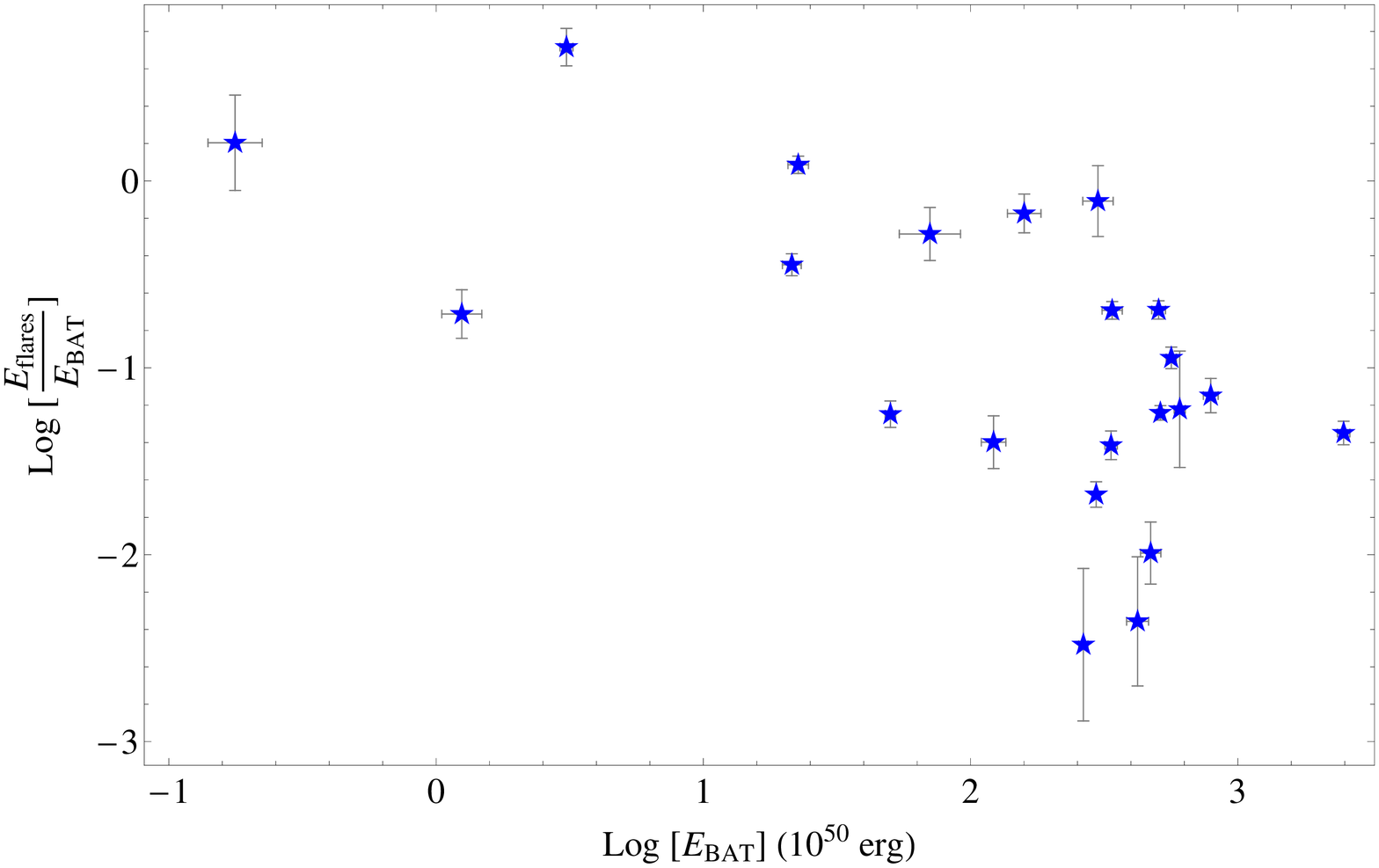}
\caption{Plot of the ratio between the total isotropic energy emitted in all the flares of a single GRB $E_{\rm flares}$ and the total isotropic energy of its prompt emission $E_{\rm BAT}$ as a function of $E_{\rm BAT}$. Although a correlation between these quantities is evident, its statistical significance is low due to the presence of two faint GRBs.}
\label{XRTBATRATIO}
\end{figure}

There is a trend for GRBs with fainter prompt emission to have also fainter flares. If we correlate the total isotropic energy emitted in all the flares of a single GRB in the XRT energy band ($E_{\rm flares}$) with the total isotropic energy of its prompt emission observed by BAT in the $15-150$ keV energy band ($E_{\rm BAT}$) we find: $E_{\rm flares}\propto E_{\rm BAT}^{0.6}$ (see Fig. \ref{XRTBATRATIO}). However the statistical significance is very low and driven considerably by two faint GRBs (GRB070724A and GRB060512). 

\subsubsection{Spectral energy distribution}\label{SED}

Except for a few cases it is very hard to measure the spectral evolution, and often the spectrum itself, of a flare because of the low SN ratio. Nevertheless, there are indications that flares have peak energies in the soft range of the X-ray spectrum, $E_{pk}< 1$ keV (Paper II). Concerning the underlying afterglow light curve, after the steep decay the spectral index does not change considerably and $\beta \sim 1$ \citep{2005astro.ph..6453C}. A different behaviour is observed during the steep decay, that is the phase during which we detect most of the flares: the spectrum can be very soft, up to $\beta \sim 3$ \citep[a beautiful example is GRB090111,][]{2009MNRAS.400L...1M}.

An indication of the Spectral Energy Distribution (SED) during the flare event can be obtained by using the $4$ sub-energy bandpasses we defined above. For each bandpass we use the $E_{\rm eff}$ introduced in Sec. \ref{shape}, computed for each GRB in the source rest frame. We define the energy density $F(E)$ in a sub-energy band as:
\begin{equation}
F(E)^i = \frac{4\pi D_l^2}{(1+z)}\frac{S^i}{E_{\rm eff}^i}\, ,
\end{equation}
where $S^i$ is the fluence of the flare\footnote{The fluence of a flare in each bandpass has been computed after the subtraction of the underlying continuum which therefore  does not contribute to the flare energy density.} in the bandpass $i$ and $D_l^2$ is the luminosity distance.

\begin{figure}
\includegraphics[width=\hsize,clip]{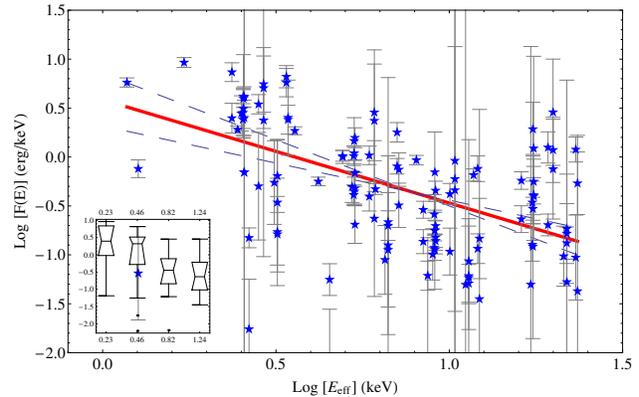}
\caption{Plot of the energy density in each band for the X-ray flares fitted in all the 4 sub-energy bandpasses (blue dots). Red solid line: best-fitting $F(E)=10^{(0.6\pm0.3)}\,E^{(-1.1\pm0.3)}$. \emph{Inset:} Boxplot of the same quantities.}
\label{Char_SED_LOG}
\end{figure}

We obtain a ``characteristic'' spectral index by fitting the energy density in each band for the X-ray flares fitted in all the $4$ sub-energy bandpasses that results to be $(-1.1\pm0.3)$ (see Fig. \ref{Char_SED_LOG}). We tentatively conclude that flares have a typical spectral index which is rather soft, although we should note that this SED is integrated over the whole duration of the flare, therefore it is dominated by the decaying part of the light curve that is generally softer. 

\begin{figure}
\includegraphics[width=\hsize,clip]{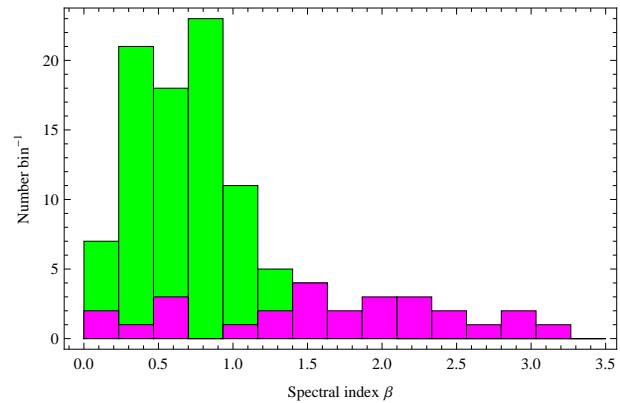}
\caption{Histogram of the spectral energy indices derived from the flare SEDs (pink rectangles) compared with the indices of the spectrum integrated over the prompt emission $T_{90}$ of the GRBs observed by BAT (green rectangles). It is evident that flares are softer, with a median $1.6$, than the prompt emission ($0.6$). The spectral indices of the BAT GRBs are reported from \citet{2008ApJS..175..179S} from December 2004 to July 2007, and from corresponding GCNs after.}
\label{BATXRTINDEX}
\end{figure}

Flares result to be much softer than the prompt emission, as expected. This is clearly shown in Fig. \ref{BATXRTINDEX} where we compare the histogram of the indices of the spectrum integrated over the prompt emission $T_{90}$ of the GRBs observed by BAT with the spectral indices of the flare SEDs derived above. The median of the flare distribution is $1.6$ with a standard deviation of $1.1$, while the median of the BAT prompt emission distribution is $0.6$ with a standard deviation of $0.4$.

\subsubsection{Hardness ratio vs. $t_{pk}$}

\begin{figure}
\includegraphics[width=\hsize,clip]{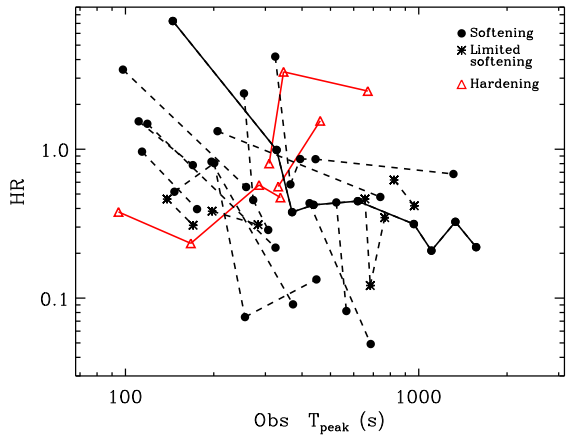}
\caption{Hardness ratio evolution with peak time for the sub-sample of $20$ GRBs
containing more than one flare and for which
an accurate measure of the profile parameters was possible. Solid
and dashed lines connect flares belonging to the same event.
Filled bullets, stars and open triangles are associated to GRBs showing
a clear softening, a limited softening or a hardening with time, respectively.
The black solid line connects the $11$ flares from GRB051117A.
A general softening trend with time is emerging.}
\label{HR_vs_tpeak_singleGRB}
\end{figure}

Fig. \ref{HR_vs_tpeak_singleGRB} demonstrates the presence of a softening
trend during the flare emission of each event. In particular, the 
hardness ratio $\rm{HR}$ is shown as a function of the observed flare 
peak time for a subsample of $20$ GRBs with more
than one fitted flare for which an accurate measure of the 
flare profiles was possible. The maximum sensitivity on the softening
is achieved computing the $\rm{HR}$ as the ratio of the counts in the 
hardest ($3-10$ keV) and in the softest ($0.3-1$ keV) energy bands as 
derived from the best fits. A general softening trend with time is 
emerging: out of $20$ events, only GRB060111A, GRB070721B and GRB080319D
show evidence of flare emission hardening with time. GRB051117A with 
11 distinct flare episodes shows instead a consistent softening from
$\sim 200$ s to $\sim 2000$ s post trigger. This leads us to conclude
that the average energy of the arriving flares becomes softer as the 
burst progresses.

The phenomena of spectral evolution during the gamma-ray prompt emission
is recognized to emerge in two distinct effects over both the entire, often
complex light curves and the individual pulses. A trend of spectral softening on the
timescale of pulses was first demonstrated by \citet{1997ApJ...486..928B}: each pulse determines a hardening during its rise time and a softening during its decay time. At the same time, a global 
spectral evolution was first suggested by \citet{1986AdSpR...6...19N} and confirmed
by \citet{1995ApJ...439..307F}. This means that, as the event progresses, the burst spectrum
tends to soften with individual, evolving pulses having impressed upon them an 
envelope that governs a global spectral decay. Fig. \ref{HR_vs_tpeak_singleGRB} extends this 
result to the flare episodes and up to thousands of seconds after the beginning of 
the explosion.

\section{Discussion}\label{discussion}

Despite many observational and theoretical efforts, the origin of the GRB outflow and the role of the magnetic field are rather uncertain. While the fireball internal shock model appears to satisfy most of the observations concerning the prompt emission, it requires various ad hoc assumptions as, for instance, the generation of a random magnetic field and efficient non-thermal particle acceleration during the collision of relativistic shells. Recent observations challenge the fireball internal shock model. \citet{2007MNRAS.376L..57K} have, e.g., pointed out that we have no strong evidence of the reverse shock that is predicted by the fireball model. The analysis carried out by \citet{2009MNRAS.395..490O} on the UVOT light curve of a large ($27$) sample of GRBs shows that the reverse shock is not the main contribution to the optical emission at early times and therefore it is not responsible for the rising optical emission. \citet{2008ApJ...687..443G} claim, however, the presence of reverse shock in GRB061126. There are, furthermore, indications that magnetic fields are rather dynamically important as shown by the polarization observed by \citet{2009Natur.462..767S}.

These observations specify the possibility that strong magnetic fields play a role during the acceleration of the outflow and the prompt emission \citep{1994MNRAS.270..480T,2001A&A...369..694S,2003astro.ph.12347L}, while any remaining magnetization at large distance can affect the interactions of the flow with the external medium. In MagnetoHydroDynamical (MHD) models for GRBs the flow is assumed to be launched by strong magnetic fields and is Poynting-flux dominated at the small distances. MHD acceleration of the jet leads to partial conversion of magnetic energy into kinetic energy in the flow. The efficiency of MHD acceleration remains uncertain and, therefore, model dependent. At large distance where the jet interacts with the external medium the jet may remain extremely magnetized \citep[as in the electromagnetic model of][]{2003astro.ph.12347L} or moderately magnetized \citep[e.g.][]{2002A&A...391.1141D,2008MNRAS.388..551T}. 

The parameter defining the magnetic content of a relativistic shell can be expressed by the ratio of the Poynting flux $F_p$ to kinetic flux $F_b$: $\sigma_\circ=F_p/F_b=B_\circ^2/(4 \pi \gamma_\circ \rho c^2)$. From this definition we have $\sigma_\circ >> 1$ for Poynting flux dominated jets while $\sigma_\circ << 1$ in the fireball model. The interaction of the flow with the circumburst medium (the stellar wind or ISM) will generally occur at a distance of $R\sim 10^{16}$ cm. It can be shown \citep[][and references therein]{2008A&A...478..747G} that for high magnetization the ejecta interact smoothly with the external medium and decelerate preventing the formation of a reverse shock. Numerical simulations have shown that for $\sigma_\circ\sim 1$, a reverse shock may still appear but it is generally weak \citep{2009A&A...494..879M}. \citet{2009Natur.462..767S} observed in GRB090102 the early afterglow optical emission with temporal slope that is characteristic of a reverse shock. This case in interesting since they detected polarization
at a $10\%$ level. In order to reconcile the presence of a large scale magnetic field and of a reverse shock, they infer a moderate magnetization of the flow of $\sigma_\circ \sim 1$. 

The present analysis of the X-ray flare properties and their comparison with the prompt emission ones should help in discriminating among different models. One fundamental question is whether the flares can originate within the internal-external shock scenario of the fireball model or they are simply due to delayed magnetic dissipation in the flow after it has powered the prompt emission \citep{2006A&A...455L...5G}.

\begin{figure}
\includegraphics[width=\hsize,clip]{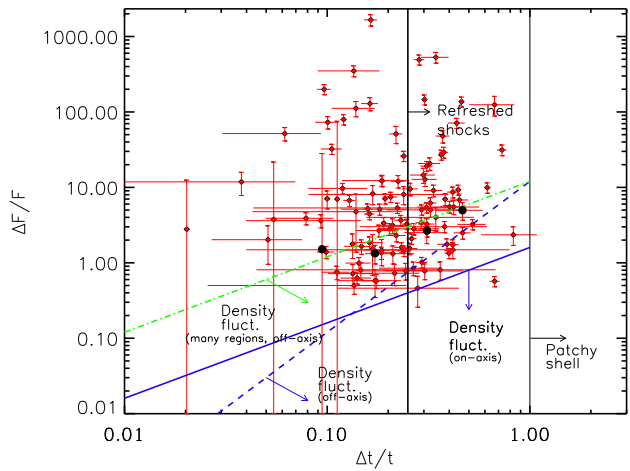}
\caption{Relative variability flux ($\Delta F/F$ ) kinematically allowed regions
as a function of relative variability time-scale $\Delta t/t$ calculated
on our sample of X-ray flares (red and black dots correspond to flare belonging to long and short GRBs respectively). The three limits shown with solid, dashed and
dot-dashed lines have been computed according to equations (7) and (A2)
of \citet{2005ApJ...631..429I}.}
\label{Ioka2005}
\end{figure}

Concerning the first possibility, we plot the variation in flux during the flare $\Delta F$ with respect to the underlying continuum $F$ versus the duration of the flare $\Delta t$ over the time of flare occurrence $t$, in analogy with the diagram presented by \citet[][, see Fig. \ref{Ioka2005}]{2005ApJ...631..429I}. We calculated the ratios $\Delta F/F$ and $\Delta t/t$ for the flare sample and plotted over the allowed regions. The two vertical lines refer respectively to bumps due to patchy shells and to bumps due to refreshed shocks. The absence of flares with $\Delta t/t \geq 1$ confirms previous results obtained in Paper I, outlining that flares cannot be due to the presence of patchy shells. On the contrary a rather large number of flares agrees with the refreshed shock limit. Only one flare in our sample may be related to on-axis density fluctuations while more flares are on the borderline that limits off-axis density fluctuations. This possibility cannot be ruled out and indeed a large amount of flares would originate from off-axis density fluctuations. These results are in agreement with what has been discussed previously in Paper I and they reveal that we cannot distinguish sharply among the sources of variability. However, a sizable fraction of the flares cannot be related to external shocks.

This is confirmed by the comparison of the present distribution of the ratio $w/t_{pk}$ with the results found by \citet{2007MNRAS.375L..46L} with the sample analysed in Paper I. In fact, the shortest timescale expected for the interaction with the circumburst medium is for adiabatic expansion in a wind environment: also in this case, however, assuming a spectral index $\beta \simeq 1$ (see Sec. \ref{SED}), the median value is $w/t_{pk}\sim 0.83$, well above the median value we found in our sample.

The indications we have seem to point toward a prolonged activity of the central engine, similar to the one producing the prompt emission. As discussed in the literature \citep{2003ApJ...592..767P,2005ApJ...630L.113K,2006ApJ...636L..29P,2006MNRAS.370L..61P,2008MNRAS.388.1729K,2009arXiv0911.1336L}, there are several mechanisms that could keep the central engine active for a long time or reactivate it in a pseudo-random way. After the formation of a black hole and accretion disk, accretion of the progenitor material left over by the collapse or disk instabilities could generate the activity we observe \citep[in some cases the instability can be strong enough to generate gravitationally bound clumps within the disk and determine disk fragmentation, see][]{2007NCimR..30..293L}. Since the accretion rate during flares is much lower than during the GRB, the neutrino mechanism is not viable for powering the flares and MHD driving is favored \citep{2008MNRAS.385L..28B}. 

One remarkable result we found in the present X-ray flares sample is that they have a ratio $w/t_{pk}$ approximately constant with time (see Fig. \ref{WTOTPK_BATSE_XRT}). This is not expected for the internal shock model since the arrival time is not related to the conditions of the collision \citep[see e.g.][]{1997ApJ...490...92K,2000ApJ...539..712R} and in fact this is not found in BATSE prompt emission pulses, where the width remains constant throughout the GRB time history \citep{1996ApJ...459..393N,2000ApJ...539..712R}. The only way to avoid this discrepancy is that it reflects a different behaviour of the progenitor generating the pulses. A positive correlation between duration and timescale in flares (see Fig. \ref{WIDTH_PEAK}), as well as an anticorrelation between duration and peak luminosity (see Fig. \ref{PeakLum_vs_tpeak}, lower panel), is expected in case of gravitational instabilities leading to fragmentation of the outer disk \citep{2006ApJ...636L..29P}. A complete explanation for the tight coupling of the evolution of the rise and decay times is, however, still missing. The analysis of the late flares sample will put more stringent constraints on this trend.

Under the assumption that the burst is produced via synchrotron emission, the internal shock model requires that the shock occurs at a distance from the center of explosion that is of the order of $10^{17}$ cm for several bursts with inferred source Lorentz factor larger than $\sim 1000$. For the deceleration radius to be larger than the internal shock radius, the circumburst medium density must be extremely low. Fast variability can be reconciled with large emitting distance if the emitting material (fundamental emitters) moves relativistically within the jet \citep{2003astro.ph.12347L,2006MNRAS.369L...5L,2009ApJ...695L..10L,2009MNRAS.395..472K}. The fundamental emitters model needs however further development. First, it is not clear how such macroscopic relativistic motions can be generated and sustained. Then, if the emitters are radiating long before their velocity points toward the observer and they continue to emit long after they move away, there is no reason for a difference the rising and decaying phase of the pulses that we found in the present sample. Furthermore, this model predicts symmetric pulses.

The asymmetry of the prompt pulses is one of the few recurrent patterns that can be distinguished among the vast range of complex GRB light-curves and the variety of GRB pulses \citep[see e.g.][and references therein]{1996ApJ...459..393N,2003ApJ...596..389K,2005ApJ...627..324N}. Studies on bright BATSE light curves revealed the so called GRB pulse paradigm, according to which narrower pulses tend to be more symmetric and have harder spectra \citep{1996ApJ...459..393N}. This paradigm has not been confirmed in the case of wide pulses \citep{2005ApJ...627..324N} or single pulses \citep{2003ApJ...596..389K}: no evidence for correlation between width and asymmetry has been found for simple FRED pulses. The finding that X-ray flares which happen hundreds of seconds after the prompt emission show asymmetry values very similar to the prompt broad pulses while being characterized by a larger width, carries important information on the GRB physical mechanism: it strongly indicates that the rise and the decay times of the pulses do not evolve independently from one another; instead, their evolution is tightly coupled. In the simplest shell collision  scenario, the pulse rise phase is due to the shell energization while the decay phase is produced by the cooling of the energized particles and the curvature of the shell. The similarity between the asymmetry values of the prompt and flare emission would point to a similar underlying physical mechanism: in this case, our results imply that the observed rate at which the shell becomes active to produce the flare emission is dependent on the decay timescale and hence the curvature of the shell. This would produce flares with a temporal profile similar to the prompt emission but stretched in time, as the time proceeds, as observed.

Flares are not necessarily the result of late central engine activity, but may be produced in the decelerating phase of the flow. High $\sigma_\circ$ flows can be prone to MHD instabilities during their interactions with the circumburst medium. The instabilities can result in energy release in localized regions through magnetic reconnection. In this picture, the presence of multiple flares, often observed in GRBs, simply corresponds to multiple reconnection regions and indeed it may be possible that dissipation in one region triggers instabilities and magnetic dissipations in nearby regions \citep{2006A&A...455L...5G}. The analysis by \citet{2006A&A...455L...5G} shows that the isotropic equivalent energy emitted in a single flare $E_{\rm flare}$ and produced by a single reconnection event is limited and related to the ratio $w/t_{pk}$ by the relation:
\begin{equation}
E_{flare}\leq 5\epsilon \left(\frac{w}{t_{pk}}\right)^3 \frac{E_{FS}}{\alpha^2}\, ,
\end{equation}
where $E_{\rm FS}$ is the isotropic energy of the forward shock, $\epsilon \sim 0.1$ the fraction of the Alfv\'en speed of the magnetic reconnection in a strongly magnetized plasma, $\alpha$ a parameter with typical values $4$ in a constant density medium and $2$ in the presence of stellar wind. The flare energy is therefore a rather strong function of the ratio $w/t_{pk}$ observed. For the XRT flares we derived $w/t_{pk} \sim 0.23$ with a standard deviation $0.14$ that would call for flares having a small fraction of the energy in the forward shock (which can be some $10$ times the observed energy in the prompt emission). The flares we observe are not in contradiction within the errors with this relation. The dimensions of the reconnection region implies that fast evolving flares are less energetic than the smoother ones. In the present sample we do not have a large enough flare range of the ratio $w/t_{pk}$ to test this effect and we find that within errors we have the same average energy. This effect will be better tested in the work in progress on late flares. It remains to be seen whether this model can quantitatively reproduce the observed decrease of the energy of the flares as function of time after the burst.

The BATSE sample reveals two types of prompt emission: prompt emission showing many overlapping short pulses (as for instance GRB991216) and prompt emission presenting a few rather wide pulses easily distinguishable and showing spectral lags. In agreement with the frequency distribution derived by \citet{2002A&A...385..377Q} the number of long lag pulses increases for GRBs detected toward the sensitivity of the BATSE threshold while bright GRBs are characterized by many short overlapping spikes (we use the words spikes and pulses as synonymous). This also implies that the distribution function of GRBs, similarly to the one of many classes of objects, increases toward the faint end and that the number of bright GRBs is very rare. The present analysis of the X-ray flares sample possibly indicates that X-ray flares belong to the class of GRBs that dominate the tail of the distribution function.

Any model put forward to explain the GRB phenomena has to account for
a some kind of memory of the central engine testified by Fig. \ref{HR_vs_tpeak_singleGRB}: 
this is able to produce distinct episodes of emission -flares- where each 
flare determines a relative hardening with respect to the previous emission, 
but with an average energy which becomes softer as the burst progresses.

\section{Summary and conclusion}

The main result of this work is that
each flare seems to retain a memory of the previous events,
so that, as time progresses, each flare is weaker and softer
than the preceding one.
In particular with this new sample of flares we confirmed all
the findings of Paper I. Furthermore, thanks to the analysis
in $4$ XRT bands and to the increased number of GRBs with redshift,
we were able to show the following:
\begin{itemize}
\item  the width of flares decreases with energy: $w \propto E^{-0.5}$,
with a power-law index that is comparable to what has been
determined for the prompt emission spikes;
\item flares are asymmetric, with a rise to decay ratio $t_{\rm rise}/t_{\rm decay}=0.49$ 
similar to prompt emission spikes.  No flare is found with $t_{\rm rise}/t_{\rm decay}>1$ ;
\item both the rise time and the decay time of each flare linearly evolve 
with time. Their evolution is such that the rise-to-decay ratio is constant with time, implying that
both time scales are stretched of the same factor;
\item the width linearly evolves with time: $w\sim 0.2\,t_{pk}$. This, together with the previous point is
one of the key feature that strongly distinguishes the flare emission with respect to the prompt phase;
\item the flare peak intensity decreases with time: on average, late time flares have lower peak intensities. However, we caution that  the relation is highly dispersed;
\item the mean isotropic flare energy of the sample is about
$10^{51}$ erg and the flare mean SED is a power-law with spectral
index $\sim 1.1$. As expected, the flares are much softer than the
prompt emission;
\item the last key point is the presence of global softening: in multiple-flare GRBs, the flares follow a softening trend which causes later time flares to be softer and softer.
\end{itemize}

The conclusion is that while we are making significant progress
in our characterization of the properties of flares and their
relation to the prompt emission spikes and prompt emission energy, at the moment there is no satisfactory model
explaining their origin, evolution and energetics.

\section*{Acknowledgments}
The anonymous referee is acknowledged for constructive criticism.
This work is supported by ASI grant SWIFT I/011/07/0, by  the
Ministry of University and Research of Italy (PRIN MIUR 2007TNYZXL), by
MAE and by the University of Milano Bicocca (Italy).

\label{lastpage}

\end{document}